\documentclass[aps,prc,reprint,groupedaddress]{revtex4-1}
\usepackage{graphicx}

\begin{document}


\title{Beta decay of $^{71,73}$Co; probing single particle states approaching doubly magic $^{78}$Ni}

\author{M. M. Rajabali$^{1,2,3}$}
\email{mrajabali@triumf.ca}
\altaffiliation[ Current affiliation:]{ TRIUMF, Vancouver BC, Canada}
\author{R. Grzywacz$^{1,4}$}
\author{S. N. Liddick$^{1,5,8}$}
\author{C. Mazzocchi$^{1,6}$}
\author{J. C. Batchelder$^{7}$}
\author{T. Baumann$^{8}$}
\author{C. R. Bingham$^{1}$}
\author{I. G. Darby$^{1,9}$}
\author{T. N. Ginter$^{8}$}
\author{S.V. Ilyushkin$^{10}$}
\author{M. Karny$^{6}$}
\author{W. Kr\'{o}las$^{3,11}$}
\author{P. F. Mantica$^{5,8}$}
\author{K. Miernik$^6$}
\author{M. Pf\"{u}tzner$^{6}$}
\author{K. P. Rykaczewski$^{4}$}
\author{D. Weisshaar$^{8}$}
\author{J. A. Winger$^{10}$}
\affiliation
  {$^{1}$University of Tennessee, Knoxville, TN 37996 USA}
\affiliation
  {$^{2}$Instituut voor Kern- en Stralingsfysica, K.U. Leuven, Celestijnenlaan 200D, B-3001 Leuven, Belgium}
\affiliation
  {$^{3}$Joint Institute for Heavy-Ion Research, Oak Ridge, TN 37831, USA}
\affiliation
  {$^{4}$Physics Division, Oak Ridge National Laboratory, Oak Ridge, TN 37831, USA}
\affiliation
  {$^{5}$Department of Chemistry, Michigan State University, East Lansing, MI 48824, USA}
\affiliation
  {$^{6}$Faculty of Physics, University of Warsaw, Warszawa, PL 00-681, Poland}
\affiliation
  {$^{7}$UNIRIB, Oak Ridge Associated Universities, Oak Ridge, TN 37831, USA}
\affiliation
  {$^{8}$National Superconducting Cyclotron Laboratory, East Lansing, MI 48824, USA}
\affiliation
  {$^{9}$Oliver Lodge Laboratory, Department of Physics, University of Liverpool, Liverpool, L69 7ZE, United Kingdom}
\affiliation
  {$^{10}$Mississippi State University, MS 39762, USA}
\affiliation
  {$^{11}$Institute of Nuclear Physics, Polish Academy of Sciences, Krak\'{o}w, PL 31-342, Poland}


\date{\today}

\begin{abstract}
Low-energy excited states in $^{71,73}$Ni populated via the $\beta$ decay of $^{71,73}$Co were investigated in an experiment performed at the National Superconducting Cyclotron Laboratory (NSCL) at Michigan State University (MSU). Detailed analysis led to the construction of level schemes of $^{71,73}$Ni, which are interpreted using systematics and analyzed using shell model calculations. The 5/2$^-$ states attributed to the the f$_{5/2}$ orbital and positive parity 5/2$^+$ and 7/2$^+$ states from the g$_{9/2}$ orbital have been identified in both $^{71,73}$Ni. In $^{71}$Ni the location of a 1/2$^{-}$ $\beta$-decaying isomer is proposed and limits are suggested as to the location of the isomer in $^{73}$Ni. The location of positive parity cluster states are also identified in $^{71,73}$Ni. Beta-delayed neutron branching ratios obtained from this data are given for both $^{71,73}$Co.
\end{abstract}


\maketitle

\section{\label{sec1}Introduction}
Nuclei in the vicinity of $^{78}$Ni have recently become accessible for experiments~\cite{granpa01,baumnature08,stolnpa04} where the persistence of nuclear magicity in neutron-rich systems can be tested. Beta decay and isomer studies using relativistic heavy ion fragmentation~\cite{berprl1991,grzprl98,muelprl99,prisprc99,sawepj04,mazplb05,mazaip05,mazepj05,hosprl05,baumnature08} were successfully carried out and the new data were used to probe the systematic variation of nuclear properties in the vicinity of $^{78}$Ni. The circumstantial evidence currently points to $^{78}$Ni being a doubly-magic nucleus with the core susceptible to polarization effects~\cite{sorprl02,lisprc04,perprl06} evidenced mainly by the properties of even-even isotopes.\par
For the odd-mass neutron-rich nickel isotopes, some information on excited states is available for $^{69-73}$Ni~\cite{mazepj05,sawepj04,grzprl98,muelprl99,prisprc99}. The most extensive work to date is on the low-lying states in $^{69}$Ni, conducted via the study of a 17/2$^{-}$ microsecond isomer~\cite{grzprl98} and through $\beta$-$\gamma$ spectroscopy of $^{69}$Co~\cite{muelprl99,prisprc99}. As a result, the level scheme for $^{69}$Ni was constructed and 1/2$^{-}$ and 5/2$^{-}$ states were identified with wavefunctions dominated by the single neutron hole configurations ($\nu$p$_{1/2}$)$^{-1}$ and ($\nu$f$_{5/2}$)$^{-1}$ respectively. Higher-spin states resulting from the angular momentum coupling of the g$_{9/2}$ neutrons and bearing similarities to those in even-even $^{70}$Ni were also observed.\par
The current work presents new experimental results for the $\beta$ decay of odd-mass $^{71}$Co and $^{73}$Co populating the low lying states in their respective decay daughters. The level scheme for  $^{71}$Ni obtained from data presented in this work has recently been published in Fig. 6 of Ref.\cite{stefprc09}, hence this work also serves to elaborate on the already published but as yet unexplained level scheme for $^{71}$Ni. Experimental evidence for level schemes developed from this data are presented and interpreted using systematics and the results of shell model calculations. These calculations are then extended to nuclei that are experimentally not available for spectroscopy in an attempt to extend the systematics of low-lying excited states to $^{75}$Ni and $^{77}$Ni.\par
In addition to the nuclear structure motivated study, $^{78}$Ni is also of interest to nuclear astrophysics~\cite{hosprl05,bosprb85}. Theoretical models of the astrophysical rapid neutron capture process (r-process)~\cite{thiephysrep93,pfenpa01,kraapJ93} predict a bottleneck around $^{78}$Ni before continuing on and ultimately leading to the creation of heavier elements. Consequently, any progress made in establishing nuclear properties near $^{78}$Ni will provide critical information to better constrain the r-process network calculations.\par
This paper begins with a description of the experimental setup in section~\ref{sec3} which is followed by the results and discussion in section~\ref{sec4} and ~\ref{discussion1} respectively.
\section{\label{sec3}Experimental setup}
The experiment was conducted at the National Superconducting Cyclotron Laboratory (NSCL) at Michigan State University (MSU). The $^{71,73}$Co isotopes studied were obtained through the fragmentation of $^{86}$Kr$^{+34}$ primary beam at an energy of 140 MeV per nucleon and an average intensity of 20 pnA impinged on a 376 mg/cm$^2$ thick $^{9}$Be target. The reaction products were separated by the A1900 fragment separator~\cite{mornimb97,mornimb03} with maximum momentum acceptance of $\Delta$p/p = 2\% determined by a mechanical slit placed at the dispersive plane. A 20 mg/cm$^2$ Kapton wedge was used at the dispersive plane to limit the number of isotopes transmitted beyond the focal plane by further increasing the A/Z separation at the focal plane of the separator. However, the wedge degrader was thin enough to give the added advantage of providing a broad range of implanted isotopes compared to previous studies in the $^{78}$Ni region~\cite{grzprl98,mazplb05,sawprc03,mazepj05,sawepj04}. Particle identification was accomplished using the standard energy loss ($\Delta$E) and time-of-flight (TOF) measurement technique~\cite{Schne1970}.\par
The implanted ions and their decay products were detected in a system of silicon detectors containing a 1 mm thick double-sided silicon strip detector (DSSD) with 16 strips on each side. The DSSD was sandwiched between two 0.5 mm thick PIN detectors. The stack of detectors was inclined at a 45$^{\circ}$ angle with respect to the beam axis to increase the effective implantation thickness. An implanted ion and its subsequent $\beta$ particle were detected in the DSSD and correlated in software. A common difficulty with such a setup is the need to detect and distinguish the signal from a high energy implanted ion and comparatively low energy $\beta$ particle. The energy range between these two events is about 1 GeV. One solution for such a detection system is to split the Si detector signal into two preamplifiers, one with high gain to observe $\beta$-decay electrons and the other with low gain for the implanted ions. A pioneering characteristic of this setup was the use of logarithmic preamplifiers~\cite{mesytec}. These preamplifiers had two output ranges: 1 keV - 10 MeV in a linear range and 10 MeV - 2 GeV in the logarithmic range. The larger analog signals from the DSSD preamplifier were scaled down to readily feed them to the digital data acquisition.\par
The $\beta$-decay events used for correlation were based on events that had
front and back strip coincidences. Once these decay events were established,
they were saved in a matrix for correlation with an ion implantation
event. The implanted ions on the other hand were detected per vertical
strip. Front and back coincidences were not possible for the ion implantation
signals due to the way the digital electronics acquisition was set up for this
experiment; the front strips recorded traces which were later analyzed to pick
out the ion events. The ion-beta correlation was therefore done using strip
correlations taking ion events that had passed the trace analysis and the
decay events that had the front back coincidence. This procedure results in a clear ion-beta correlation owing to an overall low implantation rate.\par
The digital acquisition system consisted of new generation Pixie16~\cite{xiapixie} modules made by XIA with the readout developed at the University of Tennessee and Oak Ridge National Laboratory. Each module had 16 channels with four Field Programable Gate Arrays (FPGA) and used fast Analog to Digital Conversion (ADC) at the rate of 100 mega samples per second. The modules were read in two different modes; most channels were on a self trigger, while the channels used for DSSD implantation signals were set on a triggered mode. In the latter case, the readout trigger came from the start signal for the TOF. This was necessary to avoid self-triggering on the ion-induced signals which had complicated pulse shapes due to the overloading preamplifier.\par
Gamma rays were detected using 16 high purity germanium detectors from the
Segmented Germanium Array (SeGA)~\cite{muelnima01}. These detectors were
positioned in a cylindrical long axis  geometry around the implantation point. The photopeak efficiency of this arrangement was 20.4\% at 140 keV and dropped to 7\% at 1 MeV. The signals from all silicon, scintilator and germanium detectors were read out using digital electronics as described above.
\section{\label{sec4}Results}
\begin{figure}
 \centering
 \includegraphics[totalheight=0.35\textheight]{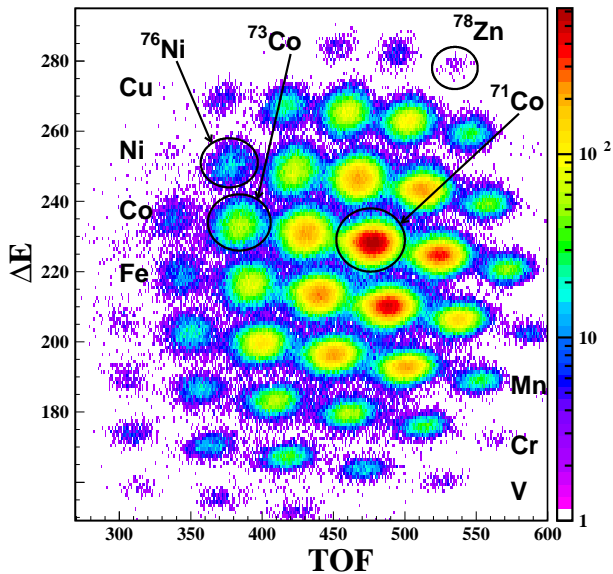}%
 \caption{\label{idplot} (Color online) The identification plot with time of flight (arbitrary units) on the horizontal axis and energy loss (arbitrary units) in a 0.5 mm PIN detector ($\Delta$E) on the vertical axis. The cobalt isotopes ($^{71,73}$Co) reported in this work and $^{76m}$Ni, and $^{78m}Zn$ used for identification have been pointed out with arrows. Rows of isotopes with the same number of protons have been labeled with their respective element symbol to give a bearing of the range of nuclei studied here.}
\end{figure}

The range of isotopes implanted in the detection setup at the focal plane of
the A1900 is shown in Fig.~\ref{idplot}. The new decay data show a marked improvement in statistics over previous studies ~\cite{mazepj05,sawepj04}. In particular, the observation of coincident $\gamma$ rays made it possible to deduce partial level schemes for isotopes of interest.\par
Gamma rays from known microsecond isomers ($^{78}$Zn, $^{76}$Ni)~\cite{mazplb05,dauplb00} were used as signatures to verify the isotope assignment of the plot in Fig.~\ref{idplot}. For each group of ions, a correlation was made between the implanted ion and its subsequent $\beta$ decay in the same DSSD strip within a selected correlation time window. On establishing the presence of the ion-$\beta$ correlation, $\gamma$ rays were then associated with the $\beta$-decaying isotope resulting in a $\beta$-$\gamma$ spectrum. Time stamping of each signal allowed for flexible time analysis based on an absolute time. The correlation time ($T_{cor}$) measured between the implantation and the subsequent decay was derived from this absolute time. For the more abundantly produced isotopes, such as $^{71,72,73}$Co,  ion-$\beta$-$\gamma$-$\gamma$ correlations (from here on referred to as $\gamma$-$\gamma$ coincidence) were extracted from the self trigger data collected for the SeGA detectors.\par
A half-life was determined for each $\beta$-decaying isotope by fitting the correlated time distribution for all $\beta$ coincident $\gamma$ rays collected. The analytical solution of Bateman equations for the case of radioactive decay in a linear chain~\cite{bat1910} was used. An additional parameter (P$_{eff}$) was introduced to account for the unknown combined efficiency of daughter ion-$\beta$-$\gamma$ decay
\begin{equation}
N_1=N_0\,e^{-\lambda_1t}
\end{equation}
\begin{equation}
 N_2=P_{eff}*N_0*\frac{\lambda_1}{\lambda_2-\lambda_1}(e^{-\lambda_1t}-e^{-\lambda_2t}),
\end{equation}
where 1 and 2 identify the parent and daughter nucleus, respectively. P$_{eff}$ remained almost constant for all nuclei fitted with a value of about 0.04. An accurate count of the number of ions that came through the delta-E detector used for the identification plot Fig.~\ref{idplot} was obscured by a malfunction of the electronic module measuring the TOF. The malfunction translated into some degree of "shadowing" on the ion identification plot shown in Fig.~\ref{idplot}. This hindered the direct measurement of the beta detection efficiency of the DSSD.\par
Time projections of the individual $\beta$-$\gamma$ peaks were also fitted to show consistency of the fit to the total $\gamma$-ray projection as well as to determine if there were any differences in lifetimes between the states. A discrepancy in the lifetimes would indicate multiple $\beta$-decaying states in the parent nucleus. Such a scenario was previously observed in $^{68}$Co and was also observed in this experiment for the decay of $^{72}$Co~\cite{mmrthesis,mmr72co}. The projections from individual peaks were fitted with a simple exponential decay with a flat baseline. This simplified procedure was justified due to the low background at long time correlations and also since $\beta$-$\gamma$ correlations should isolate single decay components.
\subsection{\label{subsection1}Beta decay of $^{71}$Co}
\begin{figure*}
\centering
\includegraphics[totalheight=0.45\textheight]{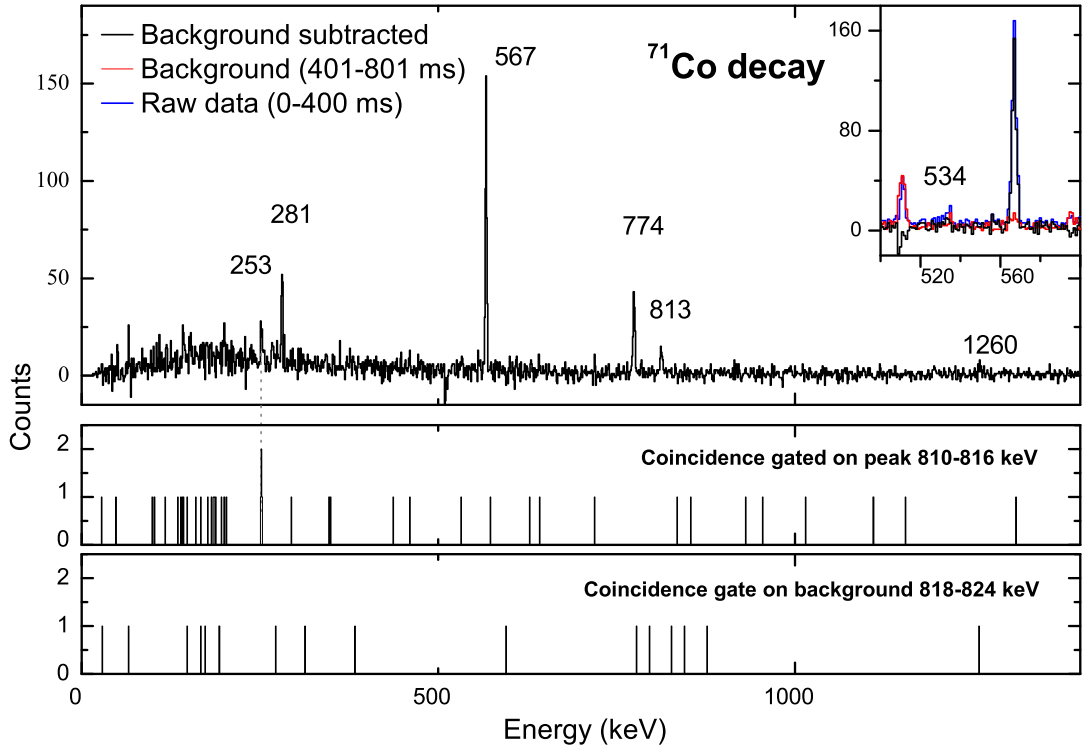}%
\caption{\label{71co_spec} (Color online)(upper panel) Three spectra are overlayed to show the background subtraction applied. The black line shows the background subtracted ion-$\beta$-gated $\gamma$-ray spectrum of $^{71}$Co. The ion-$\beta$ correlation time used was 400 ms. The 1260 keV line is from $\beta$-delayed neutron emission ($\beta_{n}$) of $^{71}$Co to $^{70}$Ni. All other labeled peaks correspond to the $\beta$ decay of $^{71}$Co to $^{71}$Ni. The blue line shows the raw spectrum and the red line is the background spectrum taken for a correlation time of 401 ms to 801 ms. The inset illustrates the presence of the 534 keV peak in the raw spectrum but it is not observed in the background subtracted spectrum. (Middle panel) A $\gamma$-$\gamma$ coincidence spectrum gated on the 813 keV peak is shown. (Lower panel) Shows a background coincidence spectrum obtained with a gate made before the 813 keV peak. The width of the background gate was the same as that of the 813 keV peak.}
\end{figure*}

The most abundantly produced isotope was $^{71}$Co, as evidenced in Fig.~\ref{idplot}. On average, 0.3 pps were implanted in the DSSD. A 400 ms time window was used for the correlation analysis. Displayed in Fig.~\ref{71co_spec} (top) is a background subtracted $\gamma$-ray spectrum associated with the $\beta$ decay of $^{71}$Co. The energies of the transitions identified in this spectrum agree with previous observations~\cite{mazepj05,sawepj04}. The $\gamma$-$\gamma$ coincidences for $^{71}$Co revealed only one cascade between the 813 keV and 253 keV transitions, as shown in Fig.~\ref{71co_spec}. The intensities of these two transitions are comparable within statistical uncertainty. Evidence for additional transitions with similar intensity was not observed in the $\gamma$-$\gamma$ coincidence spectrum. The ordering of the two transitions is interchangeable. The most intense transitions in Fig.~\ref{71co_spec} with energies at 567, 774 and 281 keV are not in coincidence. This observation presented an impediment to the construction of a more complete level scheme for $^{71}$Ni.\par
The half-life of $^{71}$Co deduced from this data is 80(3) ms, which is within the error margin of the known value of T$_{1/2}$ = 79(5) ms~\cite{mazepj05}. The fits of the time distribution for events gated on $\gamma$ rays with energy 567 keV and 774 keV gave a similar half-life (see Table~\ref{71co_gamma_energy}).\par
\begin{table}
\centering
\caption{\label{71co_gamma_energy} The energies for the prominent $\gamma$-ray peaks in $^{71}$Ni from Fig.~\ref{71co_spec} are tabulated with their relative intensity (relative to the 567 keV line).}
\begin{ruledtabular}
\begin{tabular}{c c r}
Energy (keV) & Relative intensity (\%) & T$_{1/2} (ms)$ \\
\hline \rule{0pt}{3ex}
252.6(4) & 9(2) & \\
280.8(2) & 19(4) & \\
566.9(2) & 100 & 80(6)\\
774.3(3) & 38(7) & 81(12)\\
812.8(5) & 16(3) & \\
1259.7(9) & 5(2) & $\beta_n$ \\
\end{tabular}
\end{ruledtabular}
\end{table}
Based on the structure of neighboring N=42 $^{69}$Co, the decay of cobalt
(Z=27) isotopes ought to be dominated by the Gamow-Teller transformation of a $\nu$f$_{5/2}$ neutron into a $\pi$f$_{7/2}$ proton. The ground state spin and parity of N=44 $^{71}$Co is (7/2$^{-}$), which is attributed to a $\pi$f$_{7/2}$ hole. The ground state of the N=43 daughter nucleus $^{71}$Ni is (9/2$^{+}$), dominated by an odd neutron in the $\nu$g$_{9/2}$ orbital coupled to the J=0$^+$ in the even-even $^{70}$Ni core. Allowed GT $\beta$ decay restricts the spins and parities of directly populated excited states in $^{71}$Ni to 9/2$^{-}$, 7/2$^{-}$ and 5/2$^{-}$. These simple guidelines foster the development of the level schemes for both $^{71}$Ni and $^{73}$Ni.\par
\begin{figure*}[]
\centering
\includegraphics[totalheight=0.35\textheight]{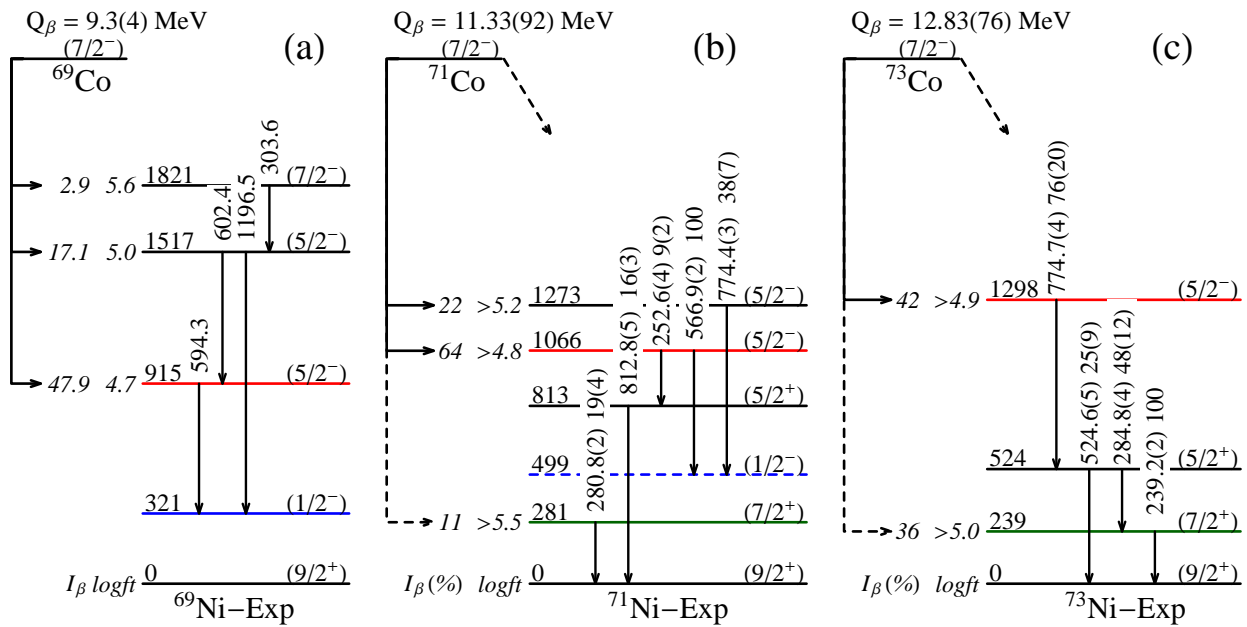}
\caption{\label{combined697173expt} (Color online) The experimental (Exp) level schemes for $^{69}$Ni, $^{71}$Ni and $^{73}$Ni. The decay energy (Q$_{\beta}$) values shown were obtained from reference~\cite{audinpa03}. (a) shows the experimental level scheme for $^{69}$Ni as obtained from Ref.~\cite{muelprl99}.
In (b), the level scheme for $^{71}$Ni is shown as proposed in this work. Feeding from GT allowed transitions populates the two 5/2$^-$ states at 1056 and 1273 keV. This newly proposed level scheme differs from an earlier, preliminary presentation in Ref.~\cite{mmr2007}. A $\beta$-decaying isomeric state is proposed at 499 keV.
(c) shows the new level scheme for $^{73}$Ni. Again the GT allowed $\beta$-decay feeds the 5/2$^-$ state at 1298 keV. The 1/2$^-$ $\beta$-decaying isomeric state observed in $^{69}$Ni and $^{71}$Ni is not observed in this nucleus.
All energy values assigned to the states and transitions are in keV.}
\end{figure*}
To interpret the new experimental data for $^{71}$Ni, a comparison to the
structure of $^{69}$Ni, shown in Fig.~\ref{combined697173expt}(a), was made since it has
been extensively studied (see~\cite{muelprl99} and references therein). There
are a few key points to note in the level scheme of $^{69}$Ni. Firstly, there
is a strong E2 transition from the 5/2$^{-}$ to the 1/2$^{-}$ state. Secondly,
there is no observed M2 transition from 5/2$^{-}$ to 9/2$^{+}$. Lastly, the
1/2$^{-}$ level is a $\beta$-decaying spin-gap isomeric state. Taking these
observations into consideration, we propose a level scheme for $^{71}$Ni which
is shown in Fig.~\ref{combined697173expt}(b),
a preliminary version of this level scheme, herein superseded, was previously
presented in Fig 6 of Ref. ~\cite{stefprc09}.\par
The most intense $\gamma$-ray transitions with energies 567, 774 and 281 keV were not in coincidence and therefore need to be placed parallel to each other in the decay scheme and parallel to the 813 keV - 253 keV cascade. Since the allowed GT decay transitions would likely populate the lowest energy 5/2$^-$ state, the 567 keV and 774 keV transitions are placed to feed the 1/2$^{-}$ level from the 5/2$^{-}$ states in analogy to the decay of $^{69}$Co (Fig.~\ref{combined697173expt}(a)).\par
In comparison to the $\beta$-decay of $^{69}$Co~\cite{muelprl99}, the 1/2$^{-}$ state is a $\beta$-decaying isomer~\cite{muelprl99,prisprc99}. The isomerism is a result of the spin difference between the 1/2$^{-}$ and 9/2$^{+}$. The transition with energy 567 keV has a half-life comparable to that of the 774 keV transition as shown in Table~\ref{71co_gamma_energy}. This suggests that the transitions must be branched out of the same $\beta$-decaying state in $^{71}$Co. One could argue that these two $\gamma$ rays, which carry significant intensity, may be transitions from 5/2$^{-}$ feeding directly to the 9/2$^{+}$. However, this possibility can be eliminated since a 5/2$^{-}$ to 9/2$^{+}$ transition would be a hindered M2 transition, hence isomeric with a single-particle lifetime estimate in the nanosecond range. After applying a hindrance factor of about 20 based on the results from the known 9/2$^{+}$ to 5/2$^{-}$ (M2) isomeric decay of $^{67}$Co~\cite{Pawlat1994,weisprc99}, the half-life is about 120 ns which is not observed in our data. The presence of these two transitions in the $\beta$-delayed $\gamma$-ray spectra in this work and in the previous ion-$\gamma$ correlation in the experiment by Sawicka et al.~\cite{sawepj04} suggests that they depopulate levels with lifetimes below the nanosecond time scale. Another possibility is a 5/2$^{-}$ to 7/2$^{+}$ E1 transition with the 7/2$^{+}$ being the first excited state (the origin of the 7/2$^{+}$ state is proposed below). This alternative can also be ruled out because the 567 and 774 keV transitions were not in coincidence with any other observed transitions in $^{71}$Ni. The probability of coincidences for all transitions assigned to $^{71}$Ni were calculated taking into account the SeGA efficiency and peak intensities from Fig.~\ref{71co_spec}. If the 567 or 774 keV transition had any coincidences with a low energy $\gamma$-ray, such evidence would have appeared as prominent peaks in the coincidence spectra.\par
Placing the 281 keV transition seen in the $\beta$-$\gamma$ spectrum was not trivial. In a separate $^{73}$Cu(1p1n,$\gamma$)$^{71}$Ni knockout reaction experiment conducted at NSCL, results from the analysis presented three prominent peaks with energy: 281, 567 and 813 keV. These knockout reaction data suggest all three transitions feed the $\beta$-decaying ground or isomeric states~\cite{stapriv}. The 281 keV is therefore assigned to be an M1 transition from the (7/2$_1^+$)\,$\rightarrow$\,(9/2$_1^+$). The spin and parity assignment rely on shell model calculations which will be presented in the discussion section. The intensity of this peak is 9\% of the total $\gamma$-ray intensity, but it cannot be fed directly from the parent nucleus via an allowed GT-transition. There are also no observed $\gamma$ rays in coincidence with the 281 keV $\gamma$ ray; therefore, no viable candidate for a 5/2$^-$\,$\rightarrow$\,7/2$^+$\,$\rightarrow$\,9/2$^+$ cascade is evident. It is assumed that the 281 keV state is fed by higher-energy transitions which were not detected by the SeGA due to the lower detection efficiencies at high energies (5\% at 1.5MeV). Another alternative is that the 7/2$^+$ state is directly fed by a forbidden transition. This point will be elaborated on in the discussion section below.\par
\begin{figure}
\centering
\includegraphics[totalheight=0.450\textheight]{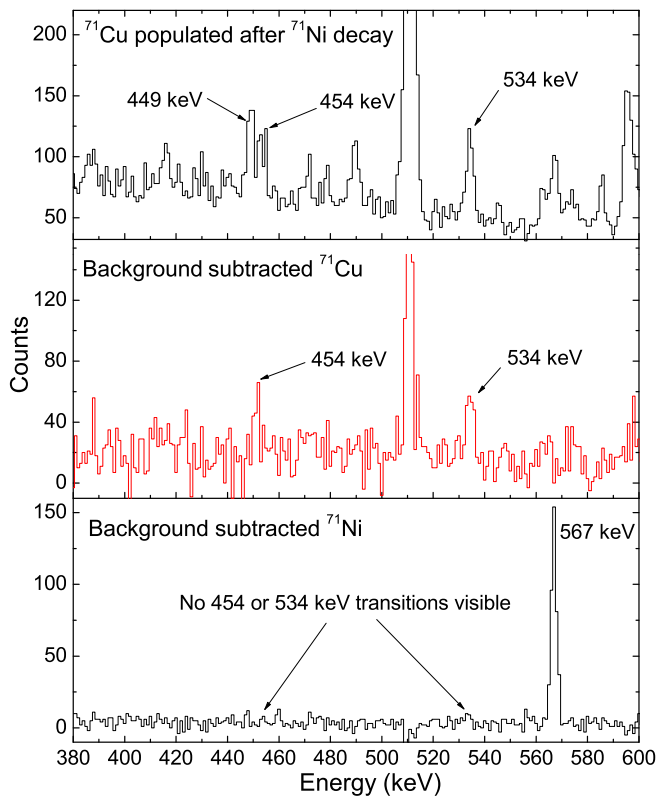}%
\caption{\label{71cu_spec} (Color online)(upper panel) The transitions populated in the granddaughter decay into $^{71}$Cu are shown with an ion-$\beta$ time correlation of 400ms to 6 seconds. The 534 and 454 keV are clearly observed although there appears to be a 449 keV background peak beside the 454 keV transition. (Middle panel) A background subtraction is attempted taking 0.4-6 s and subtracting off 3x(6-8) s. The resulting granddaughter spectrum has a peak slightly lower than 454 keV. (Lower panel) A background subtracted $^{71}$Ni spectrum identical to that in Fig.~\ref{71co_spec} is shown to compare the states populated in the daughter and the granddaughter decay.}
\end{figure}

The 813 keV $\gamma$ ray is assigned as an E2 transition in cascade with the 253 keV E1 transition. The 253 keV $\gamma$ ray from the first 5/2$^-$ state is populated directly from the 7/2$^-$ ground state of $^{71}$Co via an allowed GT transition. A 532 keV M1 transition would be expected from the 5/2$^+$ to the 7/2$^+$ state to validate the level scheme in Fig.~\ref{combined697173expt}(b). We have no clear evidence for this transition, which can be explained by two contributing factors. First, the transition would be a strongly hindered M1 transition, similar to the case of $^{51}$V~\cite{hirata1970,osnes1970,horoshko1970,zamick1971,desfes74}. The M1 hinderance comes about due to vanishing matrix elements from geometric considerations~\cite{desfes74}. In this case, the hinderance is in the order of 1000, as compared to the competing 813 keV E2 transition depopulating the same parent state. Second, the ground-state $\beta$ decay of $^{71}$Ni to $^{71}$Cu (granddaughter of $^{71}$Co) has its most intense transition at 534 keV~\cite{franprl98}. This is a strong peak observed in our data which we subtract off with the background. The inset in Fig.~\ref{71co_spec} and the spectra in Fig.~\ref{71cu_spec} show the 534 keV transition in the raw and the background spectrum which is then eliminated after the background subtraction. In conclusion, if the 532 keV was present in the data, it would be very weakly populated due to the hinderance and further obscured by the strong presence in the $^{71}$Cu granddaughter decay.\par
Inspecting the $\beta$-$\gamma$ spectra with ion-$\beta$ corelation times between 400 ms and 6 seconds, we were able to observe the 534 keV and 454 keV transitions populated in the $^{71}$Cu granddaughter. These are shown in Fig.~\ref{71cu_spec}. In this figure, it becomes clear that both the 534 and 454 keV lines are observed from the $^{71}$Ni decay since the background subtracted 0-400 ms spectrum shows no indication of these two transitions. Since the 534 keV transition in $^{71}$Cu is known to be populated via the ground state decay, it is reasonable to postulate that the 454 keV is populated via the beta-decaying isomeric state in $^{71}$Ni \cite{stefprc09}. The intensity of the 534 keV is about 41\% larger than that of the 454 keV transition (1670(232) and 980(147) counts respectively). This suggests a larger ground state feeding from $^{71}$Ni to $^{71}$Cu compared to the feeding from 1/2$^-$ $\beta$-decaying isomeric state. From the $^{71}$Co decay scheme we have deduced in this work, we conclude that there are higher lying states in $^{71}$Ni that have not been observed, which are fed from the $^{71}$Co decay and to a lesser extent some feeding to the $^{71}$Ni ground state via GT forbidden decay. Therefore the branching ratios in Fig.~\ref{combined697173expt}(b) should only be considered as the upper limit.\par
All observed $\gamma$ rays from the decay of $^{71}$Co, with their energies and corresponding intensities are listed in Table~\ref{71co_gamma_energy}. The 1260 keV transition indicated in Fig.~\ref{71co_spec} is from $^{70}$Ni after a $\beta$-delayed neutron emission. An absolute measurement of the $\beta_{n}$ branching ratio was not obtained from these data. However, using the relative intensities of the observed $\gamma$-transitions an upper limit of 2.7(9)\% was extracted.
\subsection{\label{subsection2}Beta-decay of  $^{73}$Co}
\begin{figure*}
\centering
\includegraphics[totalheight=0.45\textheight]{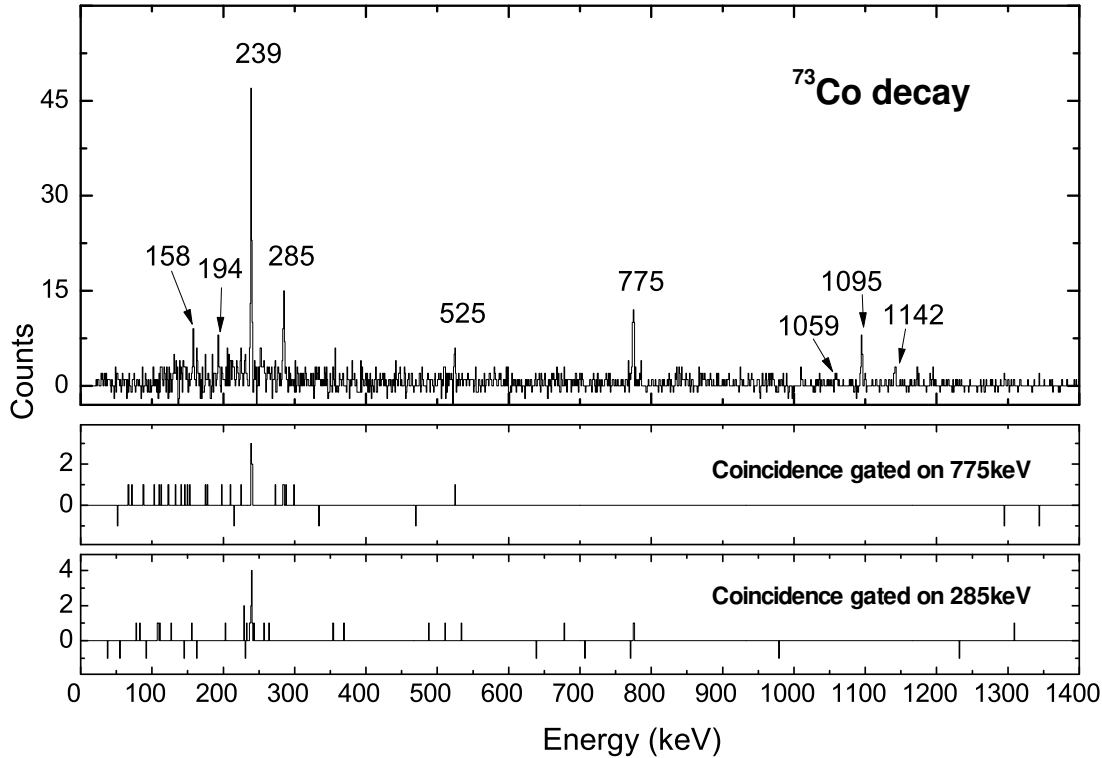}%
\caption{\label{73co_spec} (Upper panel) Gamma-ray spectrum from $\beta$-decay of $^{73}$Co. The 1095 keV line is from $\beta_{n}$ of $^{73}$Co to $^{72}$Ni, all other labeled peaks are from the $\beta$-decay of $^{73}$Co to $^{73}$Ni. (Middle panel) $\gamma$-$\gamma$ coincidence gated on the 285 keV peak and background subtracted. (Lower panel) Same as the middle but gated on the 775 keV $\gamma$-peak.}
\end{figure*}

In Fig.~\ref{73co_spec} the $\beta$-$\gamma$ spectrum, obtained from the $\beta$-decay of $^{73}$Co, is shown with a maximum ion-$\beta$ correlation time of 200 ms. The energies of the transitions identified in this spectrum mostly agree with previous observations~\cite{mazepj05,sawepj04}. The $\gamma$-ray spectrum reveals three intense transitions with energies 775 keV, 285 keV, and 239 keV. The 1095 keV transition is from $^{72}$Ni following $\beta$-delayed neutron emission. It should be noted here that in Ref.~\cite{sawepj04} the 775 keV peak observed in this work is quoted as 764 keV.\par
The half-life of $^{73}$Co decay was deduced to be 42(3) ms, which agrees with the known value of T$_{1/2}$ = 41(4) ms~\cite{mazepj05}. The half-life values obtained from fitting decay curves gated on single $\gamma$-ray transitions as well as the relative intensities of the transitions are summarized in Table~\ref{73co_gamma_energy}.\par
As is the case for $^{69,71}$Ni, the most intense transitions in $^{73}$Ni fed via the allowed GT transition should be from a 5/2${^-}$ state. The $\gamma$-$\gamma$ coincidences revealed a cascade with three
transitions: 775 keV, 285 keV, and 239 keV. This facilitated the construction
of the level scheme shown in Fig.~\ref{combined697173expt}(c). In this
level scheme, the 239 keV transition was placed to directly feed the ground
state, since it is the most intense in the cascade. The ordering of the 285
and 775 keV transitions is not evident, but a likely scenario would be to
place the 775 keV transition, which is more intense, below the 285 keV
transition. However, as can be seen in Fig.~\ref{73co_spec} (middle and lower
panel), the 525 keV transition appears in coincidence with the 775 keV
$\gamma$ ray but not with the 285 or 239 keV transitions. The 525 keV $\gamma$
ray can be caused by the summing effect of the  285 or 239 keV transitions but
this possibility can be easily eliminated. Applying summing probabilities for
the two transitions, no counts above background are expected for a 525 keV summing peak,
instead a total of 14$\pm$ 5 counts are observed in the spectrum. Thus, the four transitions can be placed as shown in Fig.~\ref{combined697173expt}(c). With this placement, the intensity of the 775 keV transition is matched within the given error by the combined intensity of the 525 and 285 keV transitions. It is presumed that there may be weak transitions from the 1298 keV or higher levels which are not observed in this data. Due to the limited information available from this experiment, it was not possible to place the 158 keV transition in the level scheme. The tentative spin and parity assignments are based on systematics from $^{69,71}$Ni and the results of shell model calculations discussed in the next section.\par
For the 5/2$^-$ to 7/2$^+$ transition there appear to be five counts at 1059 keV in the $\beta$-$\gamma$ spectrum (Fig.~\ref{73co_spec} upper panel). However, these are all within the calculated summing probability and cannot be conclusively counted as a transition. Thus, there is no evidence of a (5/2$^-$)$\rightarrow$(7/2$^+$) or a second (5/2$^+$)\,$\rightarrow$\,(7/2$^+$) transition which is consistent with the observation in $^{71}$Ni as discussed earlier. All observed $\gamma$ rays, their energies and relative intensities are listed in Table~\ref{73co_gamma_energy}. An absolute measurement of the $\beta_{n}$ branching ratio was not obtained from these data. However, an upper limit of 22(8)\% was extracted using the relative intensities of the observed 1095 keV $\gamma$ ray.
\begin{table}
\centering
\caption{\label{73co_gamma_energy} The energies of the prominent $\gamma$ rays in $^{73}$Ni from Fig.~\ref{73co_spec} (upper panel) are tabulated with their intensities relative to the most intense 239 keV line. Note the difference in value assigned to the 775 keV peak in this work and in Ref.~\cite{sawepj04}}
\begin{ruledtabular}
\begin{tabular}{c c r}
Energy (keV) & Relative intensity (\%) & T$_{1/2} (ms)$ \\
\hline \rule{0pt}{3ex}
158.0(4) & 10(4)& \\
193.5(6) & 15(5)& \\
239.2(2) & 100 & 40(9)\\
284.8(4) & 48(12) & \\
524.6(5) & 25(9) & \\
774.7(4) & 76(20) & 54(9)\\
1141.8(9) & 22(9) \\
1095(6)&39(13)&$\beta_n$\\
\end{tabular}
\end{ruledtabular}
\end{table}
\section{\label{discussion1}Discussion on odd-mass nickel isotopes}
The $\beta$-$\gamma$ spectra in Fig.~\ref{71co_spec} and Fig.~\ref{73co_spec} appear to have a pattern with two peaks in the 200 keV range and one at 775 keV. This similarity in the spectra is, however, only partially preserved in the decay schemes. Any simple structural resemblance between $^{71,73}$Ni isotopes is ruled out by the lack of coincidences in $^{71}$Ni. If the level scheme for $^{71}$Ni proposed in this work is correct, the suggested 1/2$^{-}$ state, which is fed by the most intense 567 and 774 keV transitions, should be a $\beta$-decaying isomer. Such 1/2$^{-}$$\rightarrow $9/2$^{+}$ spin gap isomers have been observed in $^{69}$Ni~\cite{prisprc99} as well as the N=50 isotones. For example, in $^{93}$Tc, which is the valence partner of $^{71}$Ni, the 1/2$^{-}$ state resides at 390 keV above the 9/2$^{+}$ ground state~\cite{Baglin1997}. In a recent article by Stefanescu et al.~\cite{stefprc09}, the authors present evidence from Coulomb excitation and $\beta$-decay experiments showing the existence of a $\beta$-decaying isomeric state in $^{71}$Ni. This is a confirmation of the 1/2$^{-}$ $\beta$-decaying isomer shown in Fig.~\ref{combined697173expt}(b).\par
\begin{figure}
\centering
\includegraphics[totalheight=0.44\textheight]{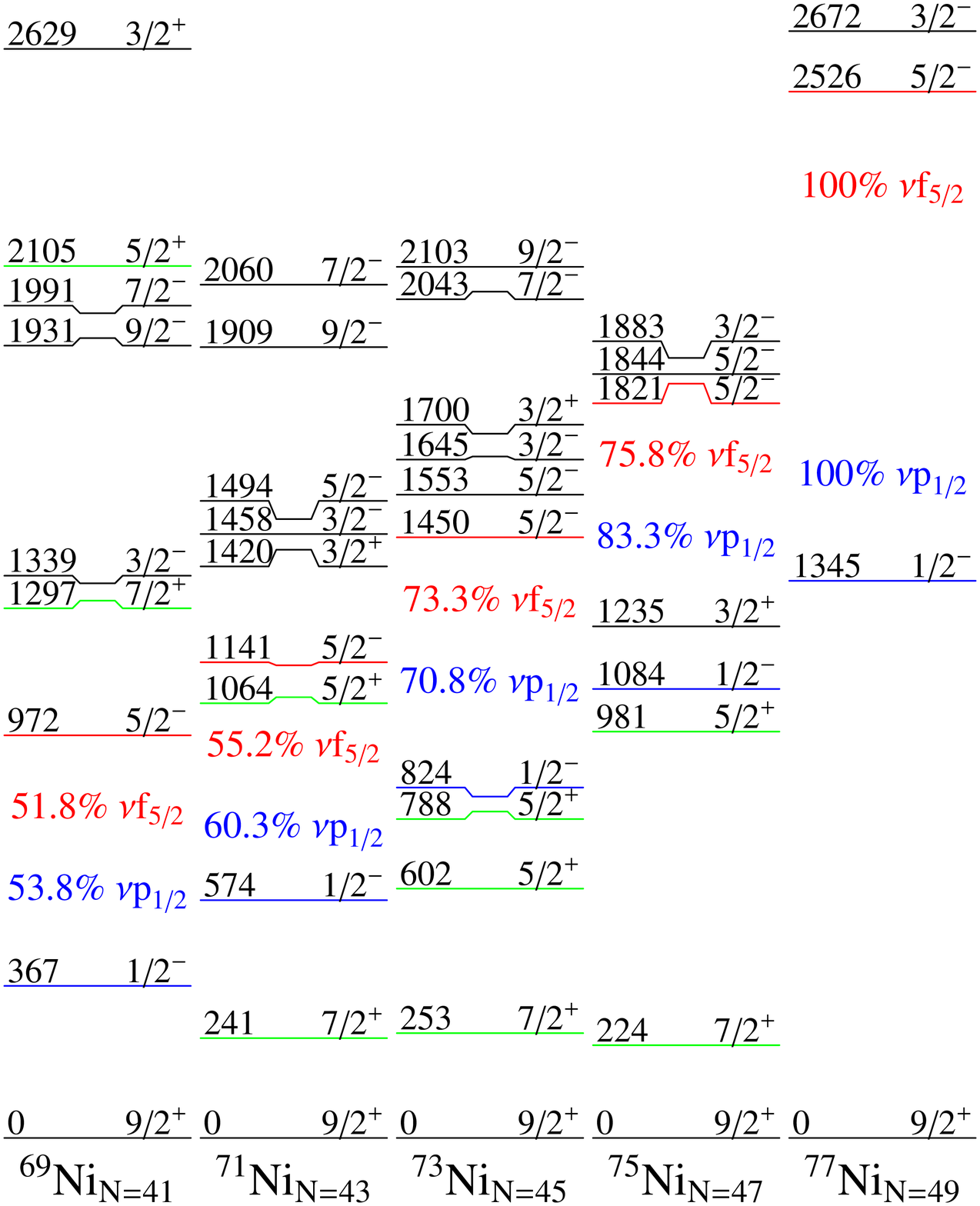}%
\caption{\label{odd_theo_ni} (Color online) Partial level schemes for $^{69-77}$Ni obtained from shell model calculations using the NR78 residual interaction~\cite{lisprc04}. The 1/2$^{-}$, 5/2$^{-}$, 7/2$^{+}$ and 9/2$^{+}$ states are all dominated by their respective single-particle wavefunctions due to $\nu$p$_{1/2}$, $\nu$f$_{5/2}$ and $\nu$g$_{9/2}$ orbitals respectively. The dominance of the single hole configurations ($\nu$p$_{1/2}$)$^{-1}$ and ($\nu$f$_{5/2}$)$^{-1}$ for the 1/2$^{-}$, 5/2$^{-}$ levels, respectively, is shown as a percentage of total wavefunction contribution for each state. All energies assigned to the respective states are in keV.}
\end{figure}

Shell model calculations have been performed using the NR78 residual interaction~\cite{lisprc04} to estimate energies of the lowest excited states in the nickel isotopes and to suggest the purity of the 5/2$^-$ and 1/2$^-$ single particle states. In these calculations, the inert core was $^{56}$Ni with the configuration space built on $\nu$p$_{3/2}$, $\nu$f$_{5/2}$, $\nu$p$_{1/2}$ and $\nu$g$_{9/2}$ orbitals and protons closed off at the f$_{7/2}$ orbital. The NR78 interaction is an empirically modified realistic
interaction. The CD-Bonn potential~\cite{cdbon96,cdbon01} was used and renormalization carried out in the G-matrix formalism. The interactions related to the (g$_{9/2})^2$ matrix elements were modified to fit experimentally known levels in the nickel isotopes. Experimental data on even-even isotopes near $^{68}$Ni~\cite{mazplb05,mmr72co,mmrthesis} are consistently interpreted with the same set of residual interactions but the effect of adjusting the TBME related to $\nu$g$_{9/2}^{2}$ J=2$^+$ had previously not been observed with the neutron rich odd mass nickel isotopes. Fig.~\ref{odd_theo_ni} shows partial level schemes for even-odd $^{69-77}$Ni obtained from these calculations. When compared to the level schemes for $^{69-73}$Ni in Fig.~\ref{combined697173expt}, very good agreement is seen between theory and experiment in the low energy structure.\par
Looking first at the negative parity states, the increase in fermi level is evidenced in Fig.~\ref{odd_theo_ni} by a steady increase in the energies of the 5/2$^{-}$ and 1/2$^{-}$ states with reference to the 9/2$^{+}$ ground state. The 5/2$^{-}$ and 1/2$^{-}$ states are dominated by ($\nu$f$_{5/2}$)$^{-1}$ and ($\nu$p$_{1/2}$)$^{-1}$ single hole configurations respectively. The contribution of the $\nu$f$_{5/2}$ and $\nu$p$_{1/2}$ orbitals obtained from these shell model calculations is given on the figure in as a percentage of the total wavefunction contribution. As would be expected, a closer inspection of the wavefunctions shows decreasing contribution of g$_{9/2}$  neutron pairs when approaching $^{77}$Ni hence a decreasing degree of configuration mixing.\par
The calculated migration of the 1/2$^{-}$ and 5/2$^{-}$ states as revealed in Fig.~\ref{odd_theo_ni} is likely responsible for bypassing the 1/2$^-$ state in $^{73}$Ni. In the case of $^{71}$Ni the intense transitions (567 and 744 keV) connect negative-parity states. The weakly-populated, parity-changing E1 transition forms a cascade (813 and 253 keV) connecting to the ground state. To explain the presently observed decay scheme in $^{73}$Ni, the 1/2$^{-}$ state needs to be placed sufficiently high in excitation energy (about 1 MeV), opening up an alternative decay path through an M2 or E3 transition to 5/2$^{+}$ and 7/2$^{+}$ states. However, the lifetime of the 1/2$^-$ state may be long enough to fall out of the narrow $\beta$-$\gamma$\,coincidence window. Thus it was not observed in the current analysis.\par
Turning now to the positive parity states, an interesting pattern is observed in Fig.~\ref{odd_theo_ni}. As the $\nu$g$_{9/2}$ orbital is filled, the positive parity states drop down in energy until midshell at $^{73}$Ni then rise again in $^{75}$Ni and $^{77}$Ni. What is most noteworthy in Fig.~\ref{odd_theo_ni} is that the positive parity states in $^{71}$Ni and $^{75}$Ni are almost at equal energies mirroring a 3 particle 3 hole effect respectively. This 3 nucleon/hole cluster effect is well explained by Paar in reference~\cite{paar_npa_1973}. Generally, coupling a few-particle cluster to the vibrational field would result in quasivibrational and quasirotational states coexisting~\cite{paar_npa_1973}. For such cases near magic nuclei, the effect results in the lowering of j-1 states (in this case the 7/2$^+$) when the particles are of the same spin (j=9/2$^+$). For the the near doubly magic cases we study in this work, the nucleus is expected to be almost spherical therefore, collectivity is at a minimum and E2 transitions strengths would be almost constant from $^{69}$Ni to $^{77}$Ni. The result is that mainly phonon multiplets will be at play. The positive parity states in $^{71}$Ni are a good display of this. Fig.~\ref{combined697173expt}(b) shows the proposed location of these states which agree well with the shell model calculation. Indeed, a closer inspection of the wavefunctions for the 7/2$^+$ and 5/2$^+$ states observed in $^{71}$Ni shows that these are mostly 3 particle cluster states with a minor contribution of other orbitals mixing in. It would be very interesting to see where these multiplets would be in $^{75}$Ni and whether the particle-hole mirror effect also holds well for the doubly magic cases. The rigidity or the softness of the doubly magic core will have a different overall effect which may be observable from the location in energy of the positive parity states in $^{75}$Ni.\par
$^{73}$Ni on the other hand shows an even stronger effect of lowering the postive parity states. From the experimental data, the location of the 7/2$^+$ and 5/2$^+$ states is proposed in Fig.~\ref{combined697173expt}(c). Results from the shell model calculation in Fig.~\ref{odd_theo_ni} suggest the existence of yet another low lying 5/2$^+$ state. Both 5/2$^+$ states have a similar $\approx$75\% contribution from ($\nu$g$_{9/2}$)$^{5}$ and an $\approx$21\% contribution from ($\nu$g$_{9/2}$)$^{-3}$. However, the majority (9\%) of ($\nu$g$_{9/2}$)$^{-3}$ contribution in 5/2$^+_1$ is from ($\nu$f$_{5/2}^5$$\nu$p$_{3/2}^4$$\nu$p$_{1/2}^1$$\nu$g$_{9/2}^7$) wavefunction while the major contribution (9\%) in 5/2$^+_2$ state is from ($\nu$f$_{5/2}^4$$\nu$p$_{3/2}^4$$\nu$p$_{1/2}^2$$\nu$g$_{9/2}^7$). It is not clear from our data which one we observe.\par
\begin{figure}
\centering
\includegraphics[totalheight=0.255\textheight]{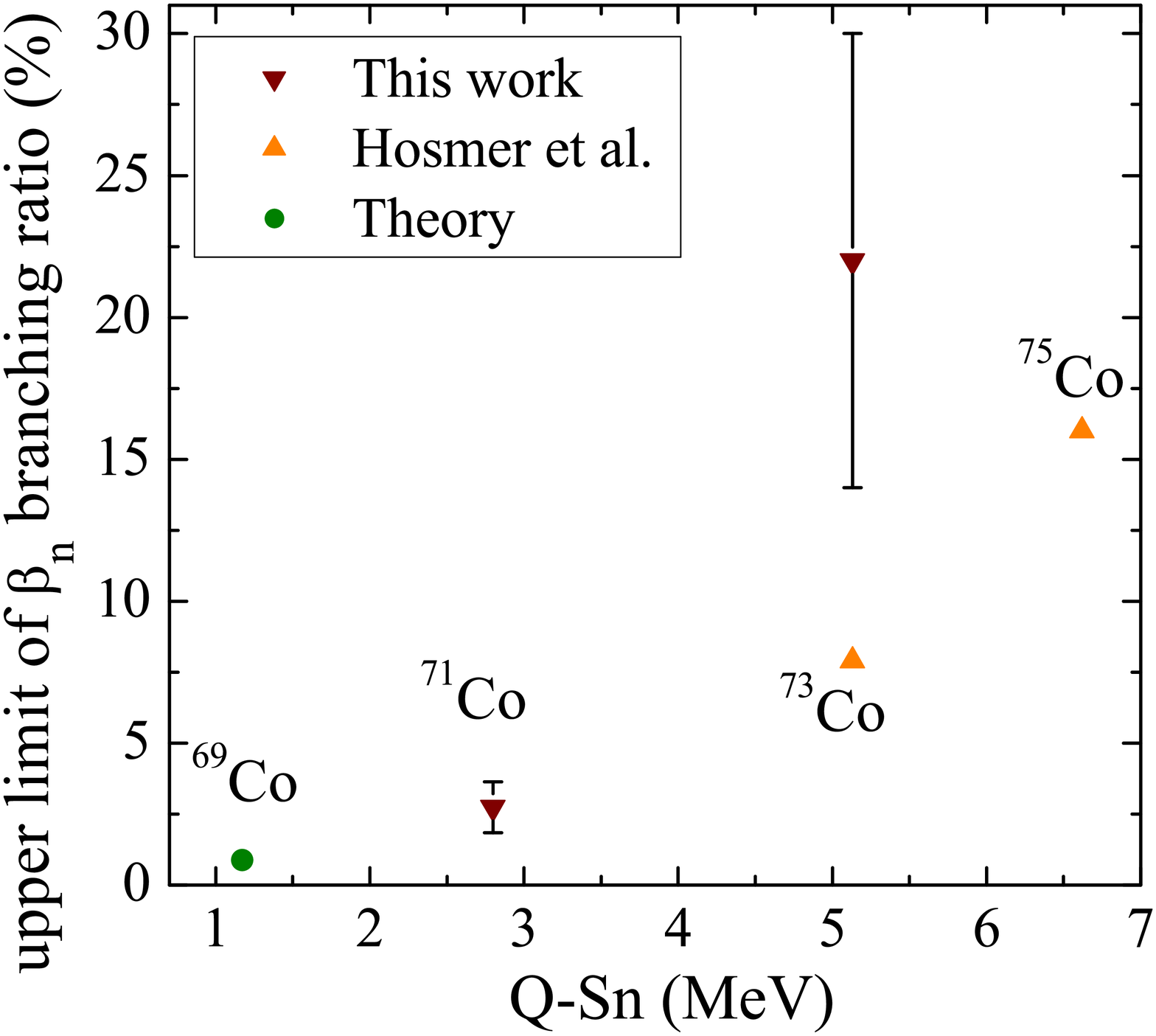}%
\caption{\label{neutron_emission} (Color online) The upper limits of $\beta{_n}$ emission branching ratios for the indicated nuclei are plotted against the difference between the $\beta$-decay Q value and neutron separation energy. The values for $^{73}$Co and $^{75}$Co (triangles) were obtained from Ref.~\cite{hosprc10} and that for $^{69}$Co (circle) was obtained from Ref.~\cite{audinpa03}.}
\end{figure}

From the $^{71}$Co and $^{73}$Co decay data, an upper limit for $\beta{_n}$ emission branching ratios of 2.7(9)\% and 22(8)\% is suggested from this work. These values were obtained based on the efficiency-corrected relative intensities of the 1260 keV and 1095 keV $\gamma$-ray transitions (see Fig.~\ref{71co_spec} and Fig.~\ref{73co_spec} respectively). This is shown in Fig.~\ref{neutron_emission} where the upper limits for the $\beta{_n}$ emission branching ratio are shown against the energy window available for each of the daughter nickel nuclei. The figure includes data from Ref.~\cite{hosprc10} for $^{73}$Co and $^{75}$Co. The upper limit for $^{73}$Co obtained from this work is grossly overestimated due to the reduced number of transitions observed in the decay of this nucleus. The lower statistics result in a large error bar.\par
The implication of increasing $\beta{_n}$ emission, together with a larger energy window for neutron emission, suggests fewer available low-lying states for an allowed GT transition and more states above the neutron separation energy. This supports the earlier conjecture that the 5/2$^{-}$ state in $^{73}$Ni resides at a relatively higher excitation energy. If a systematic increase in energy is observed for all states relative to the ground state (as for the 1/2$^{-}$ state) then it is not far-fetched to conclude that the higher lying states in $^{73}$Ni could be pushed above the neutron separation energy opening the window for a larger $\beta{_n}$ emission branching ratio.\par
With a reduction in available states for an allowed GT transition, the question that follows is the prevalence of first forbidden $\beta$ transitions. For both $^{71}$Ni and $^{73}$Ni, feeding of the 7/2$^+$ states is unaccounted for. The missing feeding is larger in $^{73}$Ni than in $^{71}$Ni. If indeed there are fewer available paths for allowed GT in the neutron bound states and a larger $\beta{_n}$ branching is observed going from $^{71}$Co to $^{73}$Co decay, then the probability of observing forbidden transitions may increase. The 7/2$^+$ state in question here has a wavefunction dominated by $\nu$ g$_{9/2}$. A possible first forbidden transition from the 7/2$^-$ ground state of $^{71,73}$Co to the the 7/2$^+$ state would involve the transformation of a neutron in $\nu$g$_{9/2}$  to a proton in the $\pi$ f$_{7/2}$. A careful study of the lifetime of the 7/2$^+$ state and the branching ratio populating it should provide more information to support the spin assignment of this state. The lifetime measurement may be extracted from the plunger experiment ~\cite{stapriv}.\par
\section{\label{sec5}Summary}
The excited 5/2$^{-}$ states in $^{71}$Ni have been placed at energies 1066 and 1273 keV. A 1/2$^{-}$ isomeric state is proposed in $^{71}$Ni at an energy of 499 keV. The position of the 5/2$_1^{-}$ state in $^{73}$Ni has tentatively been placed at 1298 keV. The 1/2$^-$ spin-gap isomer observed in the odd-mass $^{69}$Ni and $^{71}$Ni has not been identified in $^{73}$Ni. From this work it is concluded that the isomeric state, if present, must lie at an energy greater than 1 MeV allowing competing M2 and E3 transitions from the 1/2$^-$ to the low lying positive parity states.\par
The effects of significant configuration mixing is observed and is responsible for shifts of the excitation energy of low lying  5/2$^-$ and 1/2$^-$ states. These states were expected to be dominated by single particle f$_{5/2}$ and p$_{1/2}$ orbitals, but it appears that the residual interactions cause strong mixing with the neutrons in the g$_{9/2}$ orbital.\par
Positive parity cluster states have been identified in the low lying states of $^{71}$Ni and $^{73}$Ni. The overall good agreement of the systematic evolution of low-energy states from $^{69}$Ni to $^{73}$Ni with the large scale shell model calculations using NR78 interaction is critical to extending these calculation to more neutron-rich nuclei. Prediction are made for the position of excitation energies of these states in $^{75}$Ni and $^{77}$Ni approaching doubly magic $^{78}$Ni.

\begin{acknowledgments}
The author thanks the referees of PRC for useful comments. This work was supported under US DOE grants DE-AC05-00OR22725, DE-FG02-96ER40983, DE-FG02-96ER41006 and in part by the NNSA through DOE Cooperative Agreement DE-FC03-03NA00143. W. Kr\'{o}las was also supported by the Polish Ministry of Science contract No. N N202 103333. NSCL operation is supported in part by NSF grant PHY06-06007. The author was also supported by the IAP-project P6/23 of the OSTC Belgium.
\end{acknowledgments}


\begin{thebibliography}{}%
\makeatletter
\providecommand \@ifxundefined [1]{%
 \ifx #1\undefined \expandafter \@firstoftwo
 \else \expandafter \@secondoftwo
\fi
}%
\providecommand \@ifnum [1]{%
 \ifnum #1\expandafter \@firstoftwo
 \else \expandafter \@secondoftwo
\fi
}%
\providecommand \enquote [1]{``#1''}%
\providecommand \bibnamefont  [1]{#1}%
\providecommand \bibfnamefont [1]{#1}%
\providecommand \citenamefont [1]{#1}%
\providecommand\href[0]{\@sanitize\@href}%
\providecommand\@href[1]{\endgroup\@@startlink{#1}\endgroup\@@href}%
\providecommand\@@href[1]{#1\@@endlink}%
\providecommand \@sanitize [0]{\begingroup\catcode`\&12\catcode`\#12\relax}%
\@ifxundefined \pdfoutput {\@firstoftwo}{%
 \@ifnum{\z@=\pdfoutput}{\@firstoftwo}{\@secondoftwo}%
}{%
 \providecommand\@@startlink[1]{\leavevmode\special{html:<a href="#1">}}%
 \providecommand\@@endlink[0]{\special{html:</a>}}%
}{%
 \providecommand\@@startlink[1]{%
  \leavevmode
  \pdfstartlink
   attr{/Border[0 0 1 ]/H/I/C[0 1 1]}%
   user{/Subtype/Link/A<</Type/Action/S/URI/URI(#1)>>}%
  \relax
 }%
 \providecommand\@@endlink[0]{\pdfendlink}%
}%
\providecommand \url  [0]{\begingroup\@sanitize \@url }%
\providecommand \@url [1]{\endgroup\@href {#1}{\urlprefix}}%
\providecommand \urlprefix [0]{URL }%
\providecommand \Eprint[0]{\href }%
\@ifxundefined \urlstyle {%
  \providecommand \doi [1]{doi:\discretionary{}{}{}#1}%
}{%
  \providecommand \doi [0]{doi:\discretionary{}{}{}\begingroup
  \urlstyle{rm}\Url }%
}%
\providecommand \doibase [0]{http://dx.doi.org/}%
\providecommand \Doi[1]{\href{\doibase#1}}%
\providecommand \bibAnnote [3]{%
  \BibitemShut{#1}%
  \begin{quotation}\noindent
    \textsc{Key:}\ #2\\\textsc{Annotation:}\ #3%
  \end{quotation}%
}%
\providecommand \bibAnnoteFile [2]{%
  \IfFileExists{#2}{\bibAnnote {#1} {#2} {\input{#2}}}{}%
}%
\providecommand \typeout [0]{\immediate \write \m@ne }%
\providecommand \selectlanguage [0]{\@gobble}%
\providecommand \bibinfo [0]{\@secondoftwo}%
\providecommand \bibfield [0]{\@secondoftwo}%
\providecommand \translation [1]{[#1]}%
\providecommand \BibitemOpen[0]{}%
\providecommand \bibitemStop [0]{}%
\providecommand \bibitemNoStop [0]{.\EOS\space}%
\providecommand \EOS [0]{\spacefactor3000\relax}%
\providecommand \BibitemShut [1]{\csname bibitem#1\endcsname}%
\end{thebibliography}%


\begin{thebibliography}{47}%
\makeatletter
\providecommand \@ifxundefined [1]{%
 \@ifx{#1\undefined}
}%
\providecommand \@ifnum [1]{%
 \ifnum #1\expandafter \@firstoftwo
 \else \expandafter \@secondoftwo
 \fi
}%
\providecommand \@ifx [1]{%
 \ifx #1\expandafter \@firstoftwo
 \else \expandafter \@secondoftwo
 \fi
}%
\providecommand \natexlab [1]{#1}%
\providecommand \enquote  [1]{``#1''}%
\providecommand \bibnamefont  [1]{#1}%
\providecommand \bibfnamefont [1]{#1}%
\providecommand \citenamefont [1]{#1}%
\providecommand \href@noop [0]{\@secondoftwo}%
\providecommand \href [0]{\begingroup \@sanitize@url \@href}%
\providecommand \@href[1]{\@@startlink{#1}\@@href}%
\providecommand \@@href[1]{\endgroup#1\@@endlink}%
\providecommand \@sanitize@url [0]{\catcode `\\12\catcode `\$12\catcode
  `\&12\catcode `\#12\catcode `\^12\catcode `\_12\catcode `\%12\relax}%
\providecommand \@@startlink[1]{}%
\providecommand \@@endlink[0]{}%
\providecommand \url  [0]{\begingroup\@sanitize@url \@url }%
\providecommand \@url [1]{\endgroup\@href {#1}{\urlprefix }}%
\providecommand \urlprefix  [0]{URL }%
\providecommand \Eprint [0]{\href }%
\providecommand \doibase [0]{http://dx.doi.org/}%
\providecommand \selectlanguage [0]{\@gobble}%
\providecommand \bibinfo  [0]{\@secondoftwo}%
\providecommand \bibfield  [0]{\@secondoftwo}%
\providecommand \translation [1]{[#1]}%
\providecommand \BibitemOpen [0]{}%
\providecommand \bibitemStop [0]{}%
\providecommand \bibitemNoStop [0]{.\EOS\space}%
\providecommand \EOS [0]{\spacefactor3000\relax}%
\providecommand \BibitemShut  [1]{\csname bibitem#1\endcsname}%
\let\auto@bib@innerbib\@empty
\bibitem [{\citenamefont {Grawe}\ and\ \citenamefont
  {Lewitowicz}(2001)}]{granpa01}%
  \BibitemOpen
  \bibfield  {author} {\bibinfo {author} {\bibfnamefont {H.}~\bibnamefont
  {Grawe}}\ and\ \bibinfo {author} {\bibfnamefont {M.}~\bibnamefont
  {Lewitowicz}},\ }\href@noop {} {\bibfield  {journal} {\bibinfo  {journal}
  {Nuclear Physics, Section A}\ }\textbf {\bibinfo {volume} {693}},\ \bibinfo
  {pages} {116} (\bibinfo {year} {2001})}\BibitemShut {NoStop}%
\bibitem [{\citenamefont {Baumann}\ \emph {et~al.}(2007)\citenamefont
  {Baumann}, \citenamefont {Amthor}, \citenamefont {Bazin}, \citenamefont
  {Brown}, \citenamefont {Folden~III}, \citenamefont {Gade}, \citenamefont
  {Ginter}, \citenamefont {Hausmann}, \citenamefont {Mato{\v{s}}},
  \citenamefont {Morrissey} \emph {et~al.}}]{baumnature08}%
  \BibitemOpen
  \bibfield  {author} {\bibinfo {author} {\bibfnamefont {T.}~\bibnamefont
  {Baumann}}, \bibinfo {author} {\bibfnamefont {A.}~\bibnamefont {Amthor}},
  \bibinfo {author} {\bibfnamefont {D.}~\bibnamefont {Bazin}}, \bibinfo
  {author} {\bibfnamefont {B.}~\bibnamefont {Brown}}, \bibinfo {author}
  {\bibfnamefont {C.}~\bibnamefont {Folden~III}}, \bibinfo {author}
  {\bibfnamefont {A.}~\bibnamefont {Gade}}, \bibinfo {author} {\bibfnamefont
  {T.}~\bibnamefont {Ginter}}, \bibinfo {author} {\bibfnamefont
  {M.}~\bibnamefont {Hausmann}}, \bibinfo {author} {\bibfnamefont
  {M.}~\bibnamefont {Mato{\v{s}}}}, \bibinfo {author} {\bibfnamefont
  {D.}~\bibnamefont {Morrissey}},  \emph {et~al.},\ }\href@noop {} {\bibfield
  {journal} {\bibinfo  {journal} {Nature}\ }\textbf {\bibinfo {volume} {449}},\
  \bibinfo {pages} {1022} (\bibinfo {year} {2007})}\BibitemShut {NoStop}%
\bibitem [{\citenamefont {Stolz}\ \emph {et~al.}(2004)\citenamefont {Stolz},
  \citenamefont {Estrade}, \citenamefont {Davies}, \citenamefont {Ginter},
  \citenamefont {Hosmer}, \citenamefont {Kwan}, \citenamefont {Liddick},
  \citenamefont {Mantica}, \citenamefont {Mertzimekis}, \citenamefont {Montes}
  \emph {et~al.}}]{stolnpa04}%
  \BibitemOpen
  \bibfield  {author} {\bibinfo {author} {\bibfnamefont {A.}~\bibnamefont
  {Stolz}}, \bibinfo {author} {\bibfnamefont {A.}~\bibnamefont {Estrade}},
  \bibinfo {author} {\bibfnamefont {A.}~\bibnamefont {Davies}}, \bibinfo
  {author} {\bibfnamefont {T.}~\bibnamefont {Ginter}}, \bibinfo {author}
  {\bibfnamefont {P.}~\bibnamefont {Hosmer}}, \bibinfo {author} {\bibfnamefont
  {E.}~\bibnamefont {Kwan}}, \bibinfo {author} {\bibfnamefont {S.}~\bibnamefont
  {Liddick}}, \bibinfo {author} {\bibfnamefont {P.}~\bibnamefont {Mantica}},
  \bibinfo {author} {\bibfnamefont {T.}~\bibnamefont {Mertzimekis}}, \bibinfo
  {author} {\bibfnamefont {F.}~\bibnamefont {Montes}},  \emph {et~al.},\
  }\href@noop {} {\bibfield  {journal} {\bibinfo  {journal} {Nuclear Physics,
  Section A}\ }\textbf {\bibinfo {volume} {746}},\ \bibinfo {pages} {54}
  (\bibinfo {year} {2004})}\BibitemShut {NoStop}%
\bibitem [{\citenamefont {Bernas}\ \emph {et~al.}(1991)\citenamefont {Bernas},
  \citenamefont {Armbruster}, \citenamefont {Czajkowski}, \citenamefont
  {Faust}, \citenamefont {Bocquet},\ and\ \citenamefont
  {Brissot}}]{berprl1991}%
  \BibitemOpen
  \bibfield  {author} {\bibinfo {author} {\bibfnamefont {M.}~\bibnamefont
  {Bernas}}, \bibinfo {author} {\bibfnamefont {P.}~\bibnamefont {Armbruster}},
  \bibinfo {author} {\bibfnamefont {S.}~\bibnamefont {Czajkowski}}, \bibinfo
  {author} {\bibfnamefont {H.}~\bibnamefont {Faust}}, \bibinfo {author}
  {\bibfnamefont {J.}~\bibnamefont {Bocquet}}, \ and\ \bibinfo {author}
  {\bibfnamefont {R.}~\bibnamefont {Brissot}},\ }\href@noop {} {\bibfield
  {journal} {\bibinfo  {journal} {Physical review letters}\ }\textbf {\bibinfo
  {volume} {67}},\ \bibinfo {pages} {3661} (\bibinfo {year}
  {1991})}\BibitemShut {NoStop}%
\bibitem [{\citenamefont {Grzywacz}\ \emph {et~al.}(1998)\citenamefont
  {Grzywacz}, \citenamefont {B\'eraud}, \citenamefont {Borcea}, \citenamefont
  {Emsallem}, \citenamefont {Glogowski}, \citenamefont {Grawe}, \citenamefont
  {Guillemaud-Mueller}, \citenamefont {Hjorth-Jensen}, \citenamefont {Houry},
  \citenamefont {Lewitowicz}, \citenamefont {Mueller}, \citenamefont {Nowak},
  \citenamefont {P\l{}ochocki}, \citenamefont {Pf\"{u}tzner}, \citenamefont
  {Rykaczewski}, \citenamefont {Saint-Laurent}, \citenamefont {Sauvestre},
  \citenamefont {Schaefer}, \citenamefont {Sorlin}, \citenamefont {Szerypo},
  \citenamefont {Trinder}, \citenamefont {Viteritti},\ and\ \citenamefont
  {Winfield}}]{grzprl98}%
  \BibitemOpen
  \bibfield  {author} {\bibinfo {author} {\bibfnamefont {R.}~\bibnamefont
  {Grzywacz}}, \bibinfo {author} {\bibfnamefont {R.}~\bibnamefont {B\'eraud}},
  \bibinfo {author} {\bibfnamefont {C.}~\bibnamefont {Borcea}}, \bibinfo
  {author} {\bibfnamefont {A.}~\bibnamefont {Emsallem}}, \bibinfo {author}
  {\bibfnamefont {M.}~\bibnamefont {Glogowski}}, \bibinfo {author}
  {\bibfnamefont {H.}~\bibnamefont {Grawe}}, \bibinfo {author} {\bibfnamefont
  {D.}~\bibnamefont {Guillemaud-Mueller}}, \bibinfo {author} {\bibfnamefont
  {M.}~\bibnamefont {Hjorth-Jensen}}, \bibinfo {author} {\bibfnamefont
  {M.}~\bibnamefont {Houry}}, \bibinfo {author} {\bibfnamefont
  {M.}~\bibnamefont {Lewitowicz}}, \bibinfo {author} {\bibfnamefont {A.~C.}\
  \bibnamefont {Mueller}}, \bibinfo {author} {\bibfnamefont {A.}~\bibnamefont
  {Nowak}}, \bibinfo {author} {\bibfnamefont {A.}~\bibnamefont {P\l{}ochocki}},
  \bibinfo {author} {\bibfnamefont {M.}~\bibnamefont {Pf\"{u}tzner}}, \bibinfo
  {author} {\bibfnamefont {K.}~\bibnamefont {Rykaczewski}}, \bibinfo {author}
  {\bibfnamefont {M.~G.}\ \bibnamefont {Saint-Laurent}}, \bibinfo {author}
  {\bibfnamefont {J.~E.}\ \bibnamefont {Sauvestre}}, \bibinfo {author}
  {\bibfnamefont {M.}~\bibnamefont {Schaefer}}, \bibinfo {author}
  {\bibfnamefont {O.}~\bibnamefont {Sorlin}}, \bibinfo {author} {\bibfnamefont
  {J.}~\bibnamefont {Szerypo}}, \bibinfo {author} {\bibfnamefont
  {W.}~\bibnamefont {Trinder}}, \bibinfo {author} {\bibfnamefont
  {S.}~\bibnamefont {Viteritti}}, \ and\ \bibinfo {author} {\bibfnamefont
  {J.}~\bibnamefont {Winfield}},\ }\href {\doibase 10.1103/PhysRevLett.81.766}
  {\bibfield  {journal} {\bibinfo  {journal} {Physical Review Letters}\
  }\textbf {\bibinfo {volume} {81}},\ \bibinfo {pages} {766} (\bibinfo {year}
  {1998})}\BibitemShut {NoStop}%
\bibitem [{\citenamefont {Mueller}\ \emph {et~al.}(1999)\citenamefont
  {Mueller}, \citenamefont {Bruyneel}, \citenamefont {Franchoo}, \citenamefont
  {Grawe}, \citenamefont {Huyse}, \citenamefont {K{\"o}ster}, \citenamefont
  {Kratz}, \citenamefont {Kruglov}, \citenamefont {Kudryavtsev}, \citenamefont
  {Pfeiffer} \emph {et~al.}}]{muelprl99}%
  \BibitemOpen
  \bibfield  {author} {\bibinfo {author} {\bibfnamefont {W.}~\bibnamefont
  {Mueller}}, \bibinfo {author} {\bibfnamefont {B.}~\bibnamefont {Bruyneel}},
  \bibinfo {author} {\bibfnamefont {S.}~\bibnamefont {Franchoo}}, \bibinfo
  {author} {\bibfnamefont {H.}~\bibnamefont {Grawe}}, \bibinfo {author}
  {\bibfnamefont {M.}~\bibnamefont {Huyse}}, \bibinfo {author} {\bibfnamefont
  {U.}~\bibnamefont {K{\"o}ster}}, \bibinfo {author} {\bibfnamefont
  {K.}~\bibnamefont {Kratz}}, \bibinfo {author} {\bibfnamefont
  {K.}~\bibnamefont {Kruglov}}, \bibinfo {author} {\bibfnamefont
  {Y.}~\bibnamefont {Kudryavtsev}}, \bibinfo {author} {\bibfnamefont
  {B.}~\bibnamefont {Pfeiffer}},  \emph {et~al.},\ }\href@noop {} {\bibfield
  {journal} {\bibinfo  {journal} {Physical Review Letters}\ }\textbf {\bibinfo
  {volume} {83}},\ \bibinfo {pages} {3613} (\bibinfo {year}
  {1999})}\BibitemShut {NoStop}%
\bibitem [{\citenamefont {Prisciandaro}\ \emph {et~al.}(1999)\citenamefont
  {Prisciandaro}, \citenamefont {Mantica}, \citenamefont {Oros-Peusquens},
  \citenamefont {Anthony}, \citenamefont {Huhta}, \citenamefont {Lofy},\ and\
  \citenamefont {Ronningen}}]{prisprc99}%
  \BibitemOpen
  \bibfield  {author} {\bibinfo {author} {\bibfnamefont {J.~I.}\ \bibnamefont
  {Prisciandaro}}, \bibinfo {author} {\bibfnamefont {P.~F.}\ \bibnamefont
  {Mantica}}, \bibinfo {author} {\bibfnamefont {A.~M.}\ \bibnamefont
  {Oros-Peusquens}}, \bibinfo {author} {\bibfnamefont {D.~W.}\ \bibnamefont
  {Anthony}}, \bibinfo {author} {\bibfnamefont {M.}~\bibnamefont {Huhta}},
  \bibinfo {author} {\bibfnamefont {P.~A.}\ \bibnamefont {Lofy}}, \ and\
  \bibinfo {author} {\bibfnamefont {R.~M.}\ \bibnamefont {Ronningen}},\
  }\href@noop {} {\bibfield  {journal} {\bibinfo  {journal} {Physical Review
  C}\ }\textbf {\bibinfo {volume} {60}},\ \bibinfo {pages} {54307} (\bibinfo
  {year} {1999})}\BibitemShut {NoStop}%
\bibitem [{\citenamefont {Sawicka}\ \emph {et~al.}(2004)\citenamefont
  {Sawicka}, \citenamefont {Matea}, \citenamefont {Grawe}, \citenamefont
  {Grzywacz}, \citenamefont {Pf\"{u}tzner}, \citenamefont {Lewitowicz},
  \citenamefont {Daugas}, \citenamefont {Brown}, \citenamefont {Lisetskiy},
  \citenamefont {Becker}, \citenamefont {Bélier}, \citenamefont {Bingham},
  \citenamefont {Borcea}, \citenamefont {Bouchez}, \citenamefont {Buta},
  \citenamefont {Dragulescu}, \citenamefont {de~France}, \citenamefont
  {Georgiev}, \citenamefont {Giovinazzo},\ and\ \citenamefont
  {Hammache}}]{sawepj04}%
  \BibitemOpen
  \bibfield  {author} {\bibinfo {author} {\bibfnamefont {M.}~\bibnamefont
  {Sawicka}}, \bibinfo {author} {\bibfnamefont {I.}~\bibnamefont {Matea}},
  \bibinfo {author} {\bibfnamefont {H.}~\bibnamefont {Grawe}}, \bibinfo
  {author} {\bibfnamefont {R.}~\bibnamefont {Grzywacz}}, \bibinfo {author}
  {\bibfnamefont {M.}~\bibnamefont {Pf\"{u}tzner}}, \bibinfo {author}
  {\bibfnamefont {M.}~\bibnamefont {Lewitowicz}}, \bibinfo {author}
  {\bibfnamefont {J.~M.}\ \bibnamefont {Daugas}}, \bibinfo {author}
  {\bibfnamefont {B.~A.}\ \bibnamefont {Brown}}, \bibinfo {author}
  {\bibfnamefont {A.}~\bibnamefont {Lisetskiy}}, \bibinfo {author}
  {\bibfnamefont {F.}~\bibnamefont {Becker}}, \bibinfo {author} {\bibfnamefont
  {G.}~\bibnamefont {Bélier}}, \bibinfo {author} {\bibfnamefont
  {C.}~\bibnamefont {Bingham}}, \bibinfo {author} {\bibfnamefont
  {R.}~\bibnamefont {Borcea}}, \bibinfo {author} {\bibfnamefont
  {E.}~\bibnamefont {Bouchez}}, \bibinfo {author} {\bibfnamefont
  {A.}~\bibnamefont {Buta}}, \bibinfo {author} {\bibfnamefont {E.}~\bibnamefont
  {Dragulescu}}, \bibinfo {author} {\bibfnamefont {G.}~\bibnamefont
  {de~France}}, \bibinfo {author} {\bibfnamefont {G.}~\bibnamefont {Georgiev}},
  \bibinfo {author} {\bibfnamefont {J.}~\bibnamefont {Giovinazzo}}, \ and\
  \bibinfo {author} {\bibfnamefont {F.}~\bibnamefont {Hammache}},\ }\href@noop
  {} {\bibfield  {journal} {\bibinfo  {journal} {European Physical Journal A}\
  }\textbf {\bibinfo {volume} {22}},\ \bibinfo {pages} {p455 } (\bibinfo {year}
  {2004})}\BibitemShut {NoStop}%
\bibitem [{\citenamefont {Mazzocchi}\ \emph
  {et~al.}(2005{\natexlab{a}})\citenamefont {Mazzocchi}, \citenamefont
  {Grzywacz}, \citenamefont {Batchelder}, \citenamefont {Bingham},
  \citenamefont {Fong}, \citenamefont {Hamilton}, \citenamefont {Hwang},
  \citenamefont {Karny}, \citenamefont {Krolas}, \citenamefont {Liddick} \emph
  {et~al.}}]{mazplb05}%
  \BibitemOpen
  \bibfield  {author} {\bibinfo {author} {\bibfnamefont {C.}~\bibnamefont
  {Mazzocchi}}, \bibinfo {author} {\bibfnamefont {R.}~\bibnamefont {Grzywacz}},
  \bibinfo {author} {\bibfnamefont {J.}~\bibnamefont {Batchelder}}, \bibinfo
  {author} {\bibfnamefont {C.}~\bibnamefont {Bingham}}, \bibinfo {author}
  {\bibfnamefont {D.}~\bibnamefont {Fong}}, \bibinfo {author} {\bibfnamefont
  {J.}~\bibnamefont {Hamilton}}, \bibinfo {author} {\bibfnamefont
  {J.}~\bibnamefont {Hwang}}, \bibinfo {author} {\bibfnamefont
  {M.}~\bibnamefont {Karny}}, \bibinfo {author} {\bibfnamefont
  {W.}~\bibnamefont {Krolas}}, \bibinfo {author} {\bibfnamefont
  {S.}~\bibnamefont {Liddick}},  \emph {et~al.},\ }\href@noop {} {\bibfield
  {journal} {\bibinfo  {journal} {Physics Letters B}\ }\textbf {\bibinfo
  {volume} {622}},\ \bibinfo {pages} {45} (\bibinfo {year}
  {2005}{\natexlab{a}})}\BibitemShut {NoStop}%
\bibitem [{\citenamefont {Mazzocchi}\ \emph
  {et~al.}(2005{\natexlab{b}})\citenamefont {Mazzocchi}, \citenamefont
  {Grzywacz}, \citenamefont {Batchelder}, \citenamefont {Bingham},
  \citenamefont {Fong}, \citenamefont {Hamilton}, \citenamefont {Hwang},
  \citenamefont {Karny}, \citenamefont {Krolas},\ and\ \citenamefont
  {Liddick}}]{mazaip05}%
  \BibitemOpen
  \bibfield  {author} {\bibinfo {author} {\bibfnamefont {C.}~\bibnamefont
  {Mazzocchi}}, \bibinfo {author} {\bibfnamefont {R.}~\bibnamefont {Grzywacz}},
  \bibinfo {author} {\bibfnamefont {J.}~\bibnamefont {Batchelder}}, \bibinfo
  {author} {\bibfnamefont {C.}~\bibnamefont {Bingham}}, \bibinfo {author}
  {\bibfnamefont {D.}~\bibnamefont {Fong}}, \bibinfo {author} {\bibfnamefont
  {J.}~\bibnamefont {Hamilton}}, \bibinfo {author} {\bibfnamefont
  {J.}~\bibnamefont {Hwang}}, \bibinfo {author} {\bibfnamefont
  {M.}~\bibnamefont {Karny}}, \bibinfo {author} {\bibfnamefont
  {W.}~\bibnamefont {Krolas}}, \ and\ \bibinfo {author} {\bibfnamefont
  {S.}~\bibnamefont {Liddick}},\ }\href@noop {} {\bibfield  {journal} {\bibinfo
   {journal} {AIP conference proceedings}\ }\textbf {\bibinfo {volume} {764}},\
  \bibinfo {pages} {164} (\bibinfo {year} {2005}{\natexlab{b}})}\BibitemShut
  {NoStop}%
\bibitem [{\citenamefont {Mazzocchi}\ \emph
  {et~al.}(2005{\natexlab{c}})\citenamefont {Mazzocchi}, \citenamefont
  {Grzywacz}, \citenamefont {Batchelder}, \citenamefont {Bingham},
  \citenamefont {Fong}, \citenamefont {Hamilton}, \citenamefont {Hwang},
  \citenamefont {Karny}, \citenamefont {Królas}, \citenamefont {Liddick},
  \citenamefont {Morton}, \citenamefont {Mantica}, \citenamefont {Mueller},
  \citenamefont {Rykaczewski}, \citenamefont {Steiner}, \citenamefont {Stolz},\
  and\ \citenamefont {Winger}}]{mazepj05}%
  \BibitemOpen
  \bibfield  {author} {\bibinfo {author} {\bibfnamefont {C.}~\bibnamefont
  {Mazzocchi}}, \bibinfo {author} {\bibfnamefont {R.}~\bibnamefont {Grzywacz}},
  \bibinfo {author} {\bibfnamefont {J.~C.}\ \bibnamefont {Batchelder}},
  \bibinfo {author} {\bibfnamefont {C.~R.}\ \bibnamefont {Bingham}}, \bibinfo
  {author} {\bibfnamefont {D.}~\bibnamefont {Fong}}, \bibinfo {author}
  {\bibfnamefont {J.~H.}\ \bibnamefont {Hamilton}}, \bibinfo {author}
  {\bibfnamefont {J.~K.}\ \bibnamefont {Hwang}}, \bibinfo {author}
  {\bibfnamefont {M.}~\bibnamefont {Karny}}, \bibinfo {author} {\bibfnamefont
  {W.}~\bibnamefont {Królas}}, \bibinfo {author} {\bibfnamefont {S.~N.}\
  \bibnamefont {Liddick}}, \bibinfo {author} {\bibfnamefont {A.~C.}\
  \bibnamefont {Morton}}, \bibinfo {author} {\bibfnamefont {P.~F.}\
  \bibnamefont {Mantica}}, \bibinfo {author} {\bibfnamefont {W.~F.}\
  \bibnamefont {Mueller}}, \bibinfo {author} {\bibfnamefont {K.~P.}\
  \bibnamefont {Rykaczewski}}, \bibinfo {author} {\bibfnamefont
  {M.}~\bibnamefont {Steiner}}, \bibinfo {author} {\bibfnamefont
  {A.}~\bibnamefont {Stolz}}, \ and\ \bibinfo {author} {\bibfnamefont {J.~A.}\
  \bibnamefont {Winger}},\ }\href@noop {} {\bibfield  {journal} {\bibinfo
  {journal} {European Physical Journal A}\ }\textbf {\bibinfo {volume} {25}},\
  \bibinfo {pages} {p93 } (\bibinfo {year} {2005}{\natexlab{c}})}\BibitemShut
  {NoStop}%
\bibitem [{\citenamefont {Hosmer}\ \emph {et~al.}(2005)\citenamefont {Hosmer},
  \citenamefont {Schatz}, \citenamefont {Aprahamian}, \citenamefont {Arndt},
  \citenamefont {Clement}, \citenamefont {Estrade}, \citenamefont {Kratz},
  \citenamefont {Liddick}, \citenamefont {Mantica}, \citenamefont {Mueller}
  \emph {et~al.}}]{hosprl05}%
  \BibitemOpen
  \bibfield  {author} {\bibinfo {author} {\bibfnamefont {P.}~\bibnamefont
  {Hosmer}}, \bibinfo {author} {\bibfnamefont {H.}~\bibnamefont {Schatz}},
  \bibinfo {author} {\bibfnamefont {A.}~\bibnamefont {Aprahamian}}, \bibinfo
  {author} {\bibfnamefont {O.}~\bibnamefont {Arndt}}, \bibinfo {author}
  {\bibfnamefont {R.}~\bibnamefont {Clement}}, \bibinfo {author} {\bibfnamefont
  {A.}~\bibnamefont {Estrade}}, \bibinfo {author} {\bibfnamefont
  {K.}~\bibnamefont {Kratz}}, \bibinfo {author} {\bibfnamefont
  {S.}~\bibnamefont {Liddick}}, \bibinfo {author} {\bibfnamefont
  {P.}~\bibnamefont {Mantica}}, \bibinfo {author} {\bibfnamefont
  {W.}~\bibnamefont {Mueller}},  \emph {et~al.},\ }\href@noop {} {\bibfield
  {journal} {\bibinfo  {journal} {Physical Review Letters}\ }\textbf {\bibinfo
  {volume} {94}},\ \bibinfo {pages} {112501} (\bibinfo {year}
  {2005})}\BibitemShut {NoStop}%
\bibitem [{\citenamefont {Sorlin}\ \emph {et~al.}(2002)\citenamefont {Sorlin},
  \citenamefont {Leenhardt}, \citenamefont {Donzaud}, \citenamefont {Duprat},
  \citenamefont {Azaiez}, \citenamefont {Nowacki}, \citenamefont {Grawe},
  \citenamefont {Dombr\'adi}, \citenamefont {Amorini}, \citenamefont {Astier},
  \citenamefont {Baiborodin}, \citenamefont {Belleguic}, \citenamefont
  {Borcea}, \citenamefont {Bourgeois}, \citenamefont {Cullen}, \citenamefont
  {Dlouhy}, \citenamefont {Dragulescu}, \citenamefont {G\'orska}, \citenamefont
  {Gr\'evy}, \citenamefont {Guillemaud-Mueller}, \citenamefont {Hagemann},
  \citenamefont {Herskind}, \citenamefont {Kiener}, \citenamefont {Lemmon},
  \citenamefont {Lewitowicz}, \citenamefont {Lukyanov}, \citenamefont {Mayet},
  \citenamefont {deOliveiraSantos}, \citenamefont {Pantalica}, \citenamefont
  {Penionzhkevich}, \citenamefont {Pougheon}, \citenamefont {Poves},
  \citenamefont {Redon}, \citenamefont {Saint-Laurent}, \citenamefont
  {Scarpaci}, \citenamefont {Sletten}, \citenamefont {Stanoiu}, \citenamefont
  {Tarasov},\ and\ \citenamefont {Theisen}}]{sorprl02}%
  \BibitemOpen
  \bibfield  {author} {\bibinfo {author} {\bibfnamefont {O.}~\bibnamefont
  {Sorlin}}, \bibinfo {author} {\bibfnamefont {S.}~\bibnamefont {Leenhardt}},
  \bibinfo {author} {\bibfnamefont {C.}~\bibnamefont {Donzaud}}, \bibinfo
  {author} {\bibfnamefont {J.}~\bibnamefont {Duprat}}, \bibinfo {author}
  {\bibfnamefont {F.}~\bibnamefont {Azaiez}}, \bibinfo {author} {\bibfnamefont
  {F.}~\bibnamefont {Nowacki}}, \bibinfo {author} {\bibfnamefont
  {H.}~\bibnamefont {Grawe}}, \bibinfo {author} {\bibfnamefont
  {Z.}~\bibnamefont {Dombr\'adi}}, \bibinfo {author} {\bibfnamefont
  {F.}~\bibnamefont {Amorini}}, \bibinfo {author} {\bibfnamefont
  {A.}~\bibnamefont {Astier}}, \bibinfo {author} {\bibfnamefont
  {D.}~\bibnamefont {Baiborodin}}, \bibinfo {author} {\bibfnamefont
  {M.}~\bibnamefont {Belleguic}}, \bibinfo {author} {\bibfnamefont
  {C.}~\bibnamefont {Borcea}}, \bibinfo {author} {\bibfnamefont
  {C.}~\bibnamefont {Bourgeois}}, \bibinfo {author} {\bibfnamefont {D.~M.}\
  \bibnamefont {Cullen}}, \bibinfo {author} {\bibfnamefont {Z.}~\bibnamefont
  {Dlouhy}}, \bibinfo {author} {\bibfnamefont {E.}~\bibnamefont {Dragulescu}},
  \bibinfo {author} {\bibfnamefont {M.}~\bibnamefont {G\'orska}}, \bibinfo
  {author} {\bibfnamefont {S.}~\bibnamefont {Gr\'evy}}, \bibinfo {author}
  {\bibfnamefont {D.}~\bibnamefont {Guillemaud-Mueller}}, \bibinfo {author}
  {\bibfnamefont {G.}~\bibnamefont {Hagemann}}, \bibinfo {author}
  {\bibfnamefont {B.}~\bibnamefont {Herskind}}, \bibinfo {author}
  {\bibfnamefont {J.}~\bibnamefont {Kiener}}, \bibinfo {author} {\bibfnamefont
  {R.}~\bibnamefont {Lemmon}}, \bibinfo {author} {\bibfnamefont
  {M.}~\bibnamefont {Lewitowicz}}, \bibinfo {author} {\bibfnamefont {S.~M.}\
  \bibnamefont {Lukyanov}}, \bibinfo {author} {\bibfnamefont {P.}~\bibnamefont
  {Mayet}}, \bibinfo {author} {\bibfnamefont {F.}~\bibnamefont
  {deOliveiraSantos}}, \bibinfo {author} {\bibfnamefont {D.}~\bibnamefont
  {Pantalica}}, \bibinfo {author} {\bibfnamefont {Y.~E.}\ \bibnamefont
  {Penionzhkevich}}, \bibinfo {author} {\bibfnamefont {F.}~\bibnamefont
  {Pougheon}}, \bibinfo {author} {\bibfnamefont {A.}~\bibnamefont {Poves}},
  \bibinfo {author} {\bibfnamefont {N.}~\bibnamefont {Redon}}, \bibinfo
  {author} {\bibfnamefont {M.~G.}\ \bibnamefont {Saint-Laurent}}, \bibinfo
  {author} {\bibfnamefont {J.~A.}\ \bibnamefont {Scarpaci}}, \bibinfo {author}
  {\bibfnamefont {G.}~\bibnamefont {Sletten}}, \bibinfo {author} {\bibfnamefont
  {M.}~\bibnamefont {Stanoiu}}, \bibinfo {author} {\bibfnamefont
  {O.}~\bibnamefont {Tarasov}}, \ and\ \bibinfo {author} {\bibfnamefont
  {C.}~\bibnamefont {Theisen}},\ }\href {\doibase
  10.1103/PhysRevLett.88.092501} {\bibfield  {journal} {\bibinfo  {journal}
  {Phys. Rev. Lett.}\ }\textbf {\bibinfo {volume} {88}},\ \bibinfo {pages}
  {092501} (\bibinfo {year} {2002})}\BibitemShut {NoStop}%
\bibitem [{\citenamefont {Lisetskiy}\ \emph {et~al.}(2004)\citenamefont
  {Lisetskiy}, \citenamefont {Brown}, \citenamefont {Horoi},\ and\
  \citenamefont {Grawe}}]{lisprc04}%
  \BibitemOpen
  \bibfield  {author} {\bibinfo {author} {\bibfnamefont {A.~F.}\ \bibnamefont
  {Lisetskiy}}, \bibinfo {author} {\bibfnamefont {B.~A.}\ \bibnamefont
  {Brown}}, \bibinfo {author} {\bibfnamefont {M.}~\bibnamefont {Horoi}}, \ and\
  \bibinfo {author} {\bibfnamefont {H.}~\bibnamefont {Grawe}},\ }\href@noop {}
  {\bibfield  {journal} {\bibinfo  {journal} {Physical Review C}\ }\textbf
  {\bibinfo {volume} {70}},\ \bibinfo {pages} {44314} (\bibinfo {year}
  {2004})}\BibitemShut {NoStop}%
\bibitem [{\citenamefont {Perru}\ \emph {et~al.}(2006)\citenamefont {Perru},
  \citenamefont {Sorlin}, \citenamefont {Franchoo}, \citenamefont {Azaiez},
  \citenamefont {Bouchez}, \citenamefont {Bourgeois}, \citenamefont
  {Chatillon}, \citenamefont {Daugas}, \citenamefont {Dlouhy}, \citenamefont
  {Dombr{\'a}di} \emph {et~al.}}]{perprl06}%
  \BibitemOpen
  \bibfield  {author} {\bibinfo {author} {\bibfnamefont {O.}~\bibnamefont
  {Perru}}, \bibinfo {author} {\bibfnamefont {O.}~\bibnamefont {Sorlin}},
  \bibinfo {author} {\bibfnamefont {S.}~\bibnamefont {Franchoo}}, \bibinfo
  {author} {\bibfnamefont {F.}~\bibnamefont {Azaiez}}, \bibinfo {author}
  {\bibfnamefont {E.}~\bibnamefont {Bouchez}}, \bibinfo {author} {\bibfnamefont
  {C.}~\bibnamefont {Bourgeois}}, \bibinfo {author} {\bibfnamefont
  {A.}~\bibnamefont {Chatillon}}, \bibinfo {author} {\bibfnamefont
  {J.}~\bibnamefont {Daugas}}, \bibinfo {author} {\bibfnamefont
  {Z.}~\bibnamefont {Dlouhy}}, \bibinfo {author} {\bibfnamefont
  {Z.}~\bibnamefont {Dombr{\'a}di}},  \emph {et~al.},\ }\href@noop {}
  {\bibfield  {journal} {\bibinfo  {journal} {Physical Review Letters}\
  }\textbf {\bibinfo {volume} {96}},\ \bibinfo {pages} {232501} (\bibinfo
  {year} {2006})}\BibitemShut {NoStop}%
\bibitem [{\citenamefont {Stefanescu}\ \emph {et~al.}(2009)\citenamefont
  {Stefanescu}, \citenamefont {Pauwels}, \citenamefont {Bree}, \citenamefont
  {Cocolios}, \citenamefont {Diriken}, \citenamefont {Franchoo}, \citenamefont
  {Huyse}, \citenamefont {Ivanov}, \citenamefont {Kudryavtsev}, \citenamefont
  {Patronis}, \citenamefont {Walle}, \citenamefont {Duppen},\ and\
  \citenamefont {Walters}}]{stefprc09}%
  \BibitemOpen
  \bibfield  {author} {\bibinfo {author} {\bibfnamefont {I.}~\bibnamefont
  {Stefanescu}}, \bibinfo {author} {\bibfnamefont {D.}~\bibnamefont {Pauwels}},
  \bibinfo {author} {\bibfnamefont {N.}~\bibnamefont {Bree}}, \bibinfo {author}
  {\bibfnamefont {T.~E.}\ \bibnamefont {Cocolios}}, \bibinfo {author}
  {\bibfnamefont {J.}~\bibnamefont {Diriken}}, \bibinfo {author} {\bibfnamefont
  {S.}~\bibnamefont {Franchoo}}, \bibinfo {author} {\bibfnamefont
  {M.}~\bibnamefont {Huyse}}, \bibinfo {author} {\bibfnamefont
  {O.}~\bibnamefont {Ivanov}}, \bibinfo {author} {\bibfnamefont
  {Y.}~\bibnamefont {Kudryavtsev}}, \bibinfo {author} {\bibfnamefont
  {N.}~\bibnamefont {Patronis}}, \bibinfo {author} {\bibfnamefont {J.~V.~D.}\
  \bibnamefont {Walle}}, \bibinfo {author} {\bibfnamefont {P.~V.}\ \bibnamefont
  {Duppen}}, \ and\ \bibinfo {author} {\bibfnamefont {W.~B.}\ \bibnamefont
  {Walters}},\ }\href {\doibase 10.1103/PhysRevC.79.044325} {\bibfield
  {journal} {\bibinfo  {journal} {Phys. Rev. C}\ }\textbf {\bibinfo {volume}
  {79}},\ \bibinfo {pages} {044325} (\bibinfo {year} {2009})}\BibitemShut
  {NoStop}%
\bibitem [{\citenamefont {Bosch}\ \emph {et~al.}(1985)\citenamefont {Bosch},
  \citenamefont {Schmidt-Ott}, \citenamefont {Tidemand-Petersson},
  \citenamefont {Runte}, \citenamefont {Hillebrandt}, \citenamefont {Lechle},
  \citenamefont {Thielemann}, \citenamefont {Kirchner}, \citenamefont
  {Klepper}, \citenamefont {Roeckl} \emph {et~al.}}]{bosprb85}%
  \BibitemOpen
  \bibfield  {author} {\bibinfo {author} {\bibfnamefont {U.}~\bibnamefont
  {Bosch}}, \bibinfo {author} {\bibfnamefont {W.}~\bibnamefont {Schmidt-Ott}},
  \bibinfo {author} {\bibfnamefont {P.}~\bibnamefont {Tidemand-Petersson}},
  \bibinfo {author} {\bibfnamefont {E.}~\bibnamefont {Runte}}, \bibinfo
  {author} {\bibfnamefont {W.}~\bibnamefont {Hillebrandt}}, \bibinfo {author}
  {\bibfnamefont {M.}~\bibnamefont {Lechle}}, \bibinfo {author} {\bibfnamefont
  {F.}~\bibnamefont {Thielemann}}, \bibinfo {author} {\bibfnamefont
  {R.}~\bibnamefont {Kirchner}}, \bibinfo {author} {\bibfnamefont
  {O.}~\bibnamefont {Klepper}}, \bibinfo {author} {\bibfnamefont
  {E.}~\bibnamefont {Roeckl}},  \emph {et~al.},\ }\href@noop {} {\bibfield
  {journal} {\bibinfo  {journal} {Physics Letters B}\ }\textbf {\bibinfo
  {volume} {164}},\ \bibinfo {pages} {22} (\bibinfo {year} {1985})}\BibitemShut
  {NoStop}%
\bibitem [{\citenamefont {Thielemann}\ \emph {et~al.}(1993)\citenamefont
  {Thielemann}, \citenamefont {Bitouzet}, \citenamefont {Kratz}, \citenamefont
  {M{\"o}ller}, \citenamefont {Cowan},\ and\ \citenamefont
  {Truran}}]{thiephysrep93}%
  \BibitemOpen
  \bibfield  {author} {\bibinfo {author} {\bibfnamefont {F.}~\bibnamefont
  {Thielemann}}, \bibinfo {author} {\bibfnamefont {J.}~\bibnamefont
  {Bitouzet}}, \bibinfo {author} {\bibfnamefont {K.}~\bibnamefont {Kratz}},
  \bibinfo {author} {\bibfnamefont {P.}~\bibnamefont {M{\"o}ller}}, \bibinfo
  {author} {\bibfnamefont {J.}~\bibnamefont {Cowan}}, \ and\ \bibinfo {author}
  {\bibfnamefont {J.}~\bibnamefont {Truran}},\ }\href@noop {} {\bibfield
  {journal} {\bibinfo  {journal} {Physics Reports}\ }\textbf {\bibinfo {volume}
  {227}},\ \bibinfo {pages} {269} (\bibinfo {year} {1993})}\BibitemShut
  {NoStop}%
\bibitem [{\citenamefont {Pfeiffer}\ \emph {et~al.}(2001)\citenamefont
  {Pfeiffer}, \citenamefont {Kratz}, \citenamefont {Thielemann},\ and\
  \citenamefont {Walters}}]{pfenpa01}%
  \BibitemOpen
  \bibfield  {author} {\bibinfo {author} {\bibfnamefont {B.}~\bibnamefont
  {Pfeiffer}}, \bibinfo {author} {\bibfnamefont {K.}~\bibnamefont {Kratz}},
  \bibinfo {author} {\bibfnamefont {F.}~\bibnamefont {Thielemann}}, \ and\
  \bibinfo {author} {\bibfnamefont {W.}~\bibnamefont {Walters}},\ }\href@noop
  {} {\bibfield  {journal} {\bibinfo  {journal} {Nuclear Physics A}\ }\textbf
  {\bibinfo {volume} {693}},\ \bibinfo {pages} {282} (\bibinfo {year}
  {2001})}\BibitemShut {NoStop}%
\bibitem [{\citenamefont {{Kratz}}\ \emph {et~al.}(1993)\citenamefont
  {{Kratz}}, \citenamefont {{Bitouzet}}, \citenamefont {{Thielemann}},
  \citenamefont {{Moeller}},\ and\ \citenamefont {{Pfeiffer}}}]{kraapJ93}%
  \BibitemOpen
  \bibfield  {author} {\bibinfo {author} {\bibfnamefont {K.}~\bibnamefont
  {{Kratz}}}, \bibinfo {author} {\bibfnamefont {J.}~\bibnamefont {{Bitouzet}}},
  \bibinfo {author} {\bibfnamefont {F.}~\bibnamefont {{Thielemann}}}, \bibinfo
  {author} {\bibfnamefont {P.}~\bibnamefont {{Moeller}}}, \ and\ \bibinfo
  {author} {\bibnamefont {{Pfeiffer}}},\ }\href {\doibase 10.1086/172196}
  {\bibfield  {journal} {\bibinfo  {journal} {Astrophysical Journal}\ }\textbf
  {\bibinfo {volume} {403}},\ \bibinfo {pages} {216} (\bibinfo {year}
  {1993})}\BibitemShut {NoStop}%
\bibitem [{\citenamefont {Morrissey}\ and\ \citenamefont
  {Staff}(1997)}]{mornimb97}%
  \BibitemOpen
  \bibfield  {author} {\bibinfo {author} {\bibfnamefont {D.}~\bibnamefont
  {Morrissey}}\ and\ \bibinfo {author} {\bibfnamefont {N.}~\bibnamefont
  {Staff}},\ }\href@noop {} {\bibfield  {journal} {\bibinfo  {journal} {Nuclear
  Instruments and Methods in Physics Research, B}\ }\textbf {\bibinfo {volume}
  {126}},\ \bibinfo {pages} {316} (\bibinfo {year} {1997})}\BibitemShut
  {NoStop}%
\bibitem [{\citenamefont {Morrissey}\ \emph {et~al.}(2003)\citenamefont
  {Morrissey}, \citenamefont {Sherrill}, \citenamefont {Steiner}, \citenamefont
  {Stolz},\ and\ \citenamefont {Wiedenhoever}}]{mornimb03}%
  \BibitemOpen
  \bibfield  {author} {\bibinfo {author} {\bibfnamefont {D.}~\bibnamefont
  {Morrissey}}, \bibinfo {author} {\bibfnamefont {B.}~\bibnamefont {Sherrill}},
  \bibinfo {author} {\bibfnamefont {M.}~\bibnamefont {Steiner}}, \bibinfo
  {author} {\bibfnamefont {A.}~\bibnamefont {Stolz}}, \ and\ \bibinfo {author}
  {\bibfnamefont {I.}~\bibnamefont {Wiedenhoever}},\ }\href@noop {} {\bibfield
  {journal} {\bibinfo  {journal} {Nuclear Instruments and Methods in Physics
  Research, B}\ }\textbf {\bibinfo {volume} {204}},\ \bibinfo {pages} {90}
  (\bibinfo {year} {2003})}\BibitemShut {NoStop}%
\bibitem [{\citenamefont {Sawicka}\ \emph {et~al.}(2003)\citenamefont
  {Sawicka}, \citenamefont {Grzywacz}, \citenamefont {Matea}, \citenamefont
  {Grawe}, \citenamefont {Pf\"utzner}, \citenamefont {Daugas}, \citenamefont
  {Lewitowicz}, \citenamefont {Balabanski}, \citenamefont {Becker},
  \citenamefont {B\'elier}, \citenamefont {Bingham}, \citenamefont {Borcea},
  \citenamefont {Bouchez}, \citenamefont {Buta}, \citenamefont {La~Commara},
  \citenamefont {Dragulescu}, \citenamefont {de~France}, \citenamefont
  {Georgiev}, \citenamefont {Giovinazzo}, \citenamefont {G\'orska},
  \citenamefont {Hammache}, \citenamefont {Hass}, \citenamefont {Hellstr\"om},
  \citenamefont {Ibrahim}, \citenamefont {Janas}, \citenamefont {Mach},
  \citenamefont {Mayet}, \citenamefont {M\'eot}, \citenamefont {Negoita},
  \citenamefont {Neyens}, \citenamefont {de~Oliveira~Santos}, \citenamefont
  {Page}, \citenamefont {Perru}, \citenamefont {Podoly\'ak}, \citenamefont
  {Roig}, \citenamefont {Rykaczewski}, \citenamefont {Saint-Laurent},
  \citenamefont {Sauvestre}, \citenamefont {Sorlin}, \citenamefont {Stanoiu},
  \citenamefont {Stefan}, \citenamefont {Stodel}, \citenamefont {Theisen},
  \citenamefont {Verney},\ and\ \citenamefont {Zylicz}}]{sawprc03}%
  \BibitemOpen
  \bibfield  {author} {\bibinfo {author} {\bibfnamefont {M.}~\bibnamefont
  {Sawicka}}, \bibinfo {author} {\bibfnamefont {R.}~\bibnamefont {Grzywacz}},
  \bibinfo {author} {\bibfnamefont {I.}~\bibnamefont {Matea}}, \bibinfo
  {author} {\bibfnamefont {H.}~\bibnamefont {Grawe}}, \bibinfo {author}
  {\bibfnamefont {M.}~\bibnamefont {Pf\"utzner}}, \bibinfo {author}
  {\bibfnamefont {J.~M.}\ \bibnamefont {Daugas}}, \bibinfo {author}
  {\bibfnamefont {M.}~\bibnamefont {Lewitowicz}}, \bibinfo {author}
  {\bibfnamefont {D.~L.}\ \bibnamefont {Balabanski}}, \bibinfo {author}
  {\bibfnamefont {F.}~\bibnamefont {Becker}}, \bibinfo {author} {\bibfnamefont
  {G.}~\bibnamefont {B\'elier}}, \bibinfo {author} {\bibfnamefont
  {C.}~\bibnamefont {Bingham}}, \bibinfo {author} {\bibfnamefont
  {C.}~\bibnamefont {Borcea}}, \bibinfo {author} {\bibfnamefont
  {E.}~\bibnamefont {Bouchez}}, \bibinfo {author} {\bibfnamefont
  {A.}~\bibnamefont {Buta}}, \bibinfo {author} {\bibfnamefont {M.}~\bibnamefont
  {La~Commara}}, \bibinfo {author} {\bibfnamefont {E.}~\bibnamefont
  {Dragulescu}}, \bibinfo {author} {\bibfnamefont {G.}~\bibnamefont
  {de~France}}, \bibinfo {author} {\bibfnamefont {G.}~\bibnamefont {Georgiev}},
  \bibinfo {author} {\bibfnamefont {J.}~\bibnamefont {Giovinazzo}}, \bibinfo
  {author} {\bibfnamefont {M.}~\bibnamefont {G\'orska}}, \bibinfo {author}
  {\bibfnamefont {F.}~\bibnamefont {Hammache}}, \bibinfo {author}
  {\bibfnamefont {M.}~\bibnamefont {Hass}}, \bibinfo {author} {\bibfnamefont
  {M.}~\bibnamefont {Hellstr\"om}}, \bibinfo {author} {\bibfnamefont
  {F.}~\bibnamefont {Ibrahim}}, \bibinfo {author} {\bibfnamefont
  {Z.}~\bibnamefont {Janas}}, \bibinfo {author} {\bibfnamefont
  {H.}~\bibnamefont {Mach}}, \bibinfo {author} {\bibfnamefont {P.}~\bibnamefont
  {Mayet}}, \bibinfo {author} {\bibfnamefont {V.}~\bibnamefont {M\'eot}},
  \bibinfo {author} {\bibfnamefont {F.}~\bibnamefont {Negoita}}, \bibinfo
  {author} {\bibfnamefont {G.}~\bibnamefont {Neyens}}, \bibinfo {author}
  {\bibfnamefont {F.}~\bibnamefont {de~Oliveira~Santos}}, \bibinfo {author}
  {\bibfnamefont {R.~D.}\ \bibnamefont {Page}}, \bibinfo {author}
  {\bibfnamefont {O.}~\bibnamefont {Perru}}, \bibinfo {author} {\bibfnamefont
  {Z.}~\bibnamefont {Podoly\'ak}}, \bibinfo {author} {\bibfnamefont
  {O.}~\bibnamefont {Roig}}, \bibinfo {author} {\bibfnamefont {K.~P.}\
  \bibnamefont {Rykaczewski}}, \bibinfo {author} {\bibfnamefont {M.~G.}\
  \bibnamefont {Saint-Laurent}}, \bibinfo {author} {\bibfnamefont {J.~E.}\
  \bibnamefont {Sauvestre}}, \bibinfo {author} {\bibfnamefont {O.}~\bibnamefont
  {Sorlin}}, \bibinfo {author} {\bibfnamefont {M.}~\bibnamefont {Stanoiu}},
  \bibinfo {author} {\bibfnamefont {I.}~\bibnamefont {Stefan}}, \bibinfo
  {author} {\bibfnamefont {C.}~\bibnamefont {Stodel}}, \bibinfo {author}
  {\bibfnamefont {C.}~\bibnamefont {Theisen}}, \bibinfo {author} {\bibfnamefont
  {D.}~\bibnamefont {Verney}}, \ and\ \bibinfo {author} {\bibfnamefont
  {J.}~\bibnamefont {Zylicz}},\ }\href {\doibase 10.1103/PhysRevC.68.044304}
  {\bibfield  {journal} {\bibinfo  {journal} {Phys. Rev. C}\ }\textbf {\bibinfo
  {volume} {68}},\ \bibinfo {pages} {044304} (\bibinfo {year}
  {2003})}\BibitemShut {NoStop}%
\bibitem [{\citenamefont {Schneider}\ \emph {et~al.}(1970)\citenamefont
  {Schneider}, \citenamefont {Kohlmeyer},\ and\ \citenamefont
  {Bock}}]{Schne1970}%
  \BibitemOpen
  \bibfield  {author} {\bibinfo {author} {\bibfnamefont {W.}~\bibnamefont
  {Schneider}}, \bibinfo {author} {\bibfnamefont {B.}~\bibnamefont
  {Kohlmeyer}}, \ and\ \bibinfo {author} {\bibfnamefont {R.}~\bibnamefont
  {Bock}},\ }\href {\doibase DOI: 10.1016/0029-554X(70)90212-0} {\bibfield
  {journal} {\bibinfo  {journal} {Nuclear Instruments and Methods}\ }\textbf
  {\bibinfo {volume} {87}},\ \bibinfo {pages} {253 } (\bibinfo {year}
  {1970})}\BibitemShut {NoStop}%
\bibitem [{\citenamefont {mesytec GmbH \& Co.~KG}(2006)}]{mesytec}%
  \BibitemOpen
  \bibfield  {author} {\bibinfo {author} {\bibnamefont {mesytec GmbH \&
  Co.~KG}},\ }\href@noop {} {\enquote {\bibinfo {title} {{Multichannel
  Logarithmic preamplifier}},}\ } (\bibinfo {year} {2006}),\ \bibinfo {note}
  {http://www.mesytec.com}\BibitemShut {NoStop}%
\bibitem [{\citenamefont {Hennig}\ \emph {et~al.}(2007)\citenamefont {Hennig},
  \citenamefont {Tan}, \citenamefont {Walby}, \citenamefont {Grudberg},
  \citenamefont {Fallu-Labruyere}, \citenamefont {Warburton}, \citenamefont
  {Vaman}, \citenamefont {Starosta},\ and\ \citenamefont {Miller}}]{xiapixie}%
  \BibitemOpen
  \bibfield  {author} {\bibinfo {author} {\bibfnamefont {W.}~\bibnamefont
  {Hennig}}, \bibinfo {author} {\bibfnamefont {H.}~\bibnamefont {Tan}},
  \bibinfo {author} {\bibfnamefont {M.}~\bibnamefont {Walby}}, \bibinfo
  {author} {\bibfnamefont {P.}~\bibnamefont {Grudberg}}, \bibinfo {author}
  {\bibfnamefont {A.}~\bibnamefont {Fallu-Labruyere}}, \bibinfo {author}
  {\bibfnamefont {W.}~\bibnamefont {Warburton}}, \bibinfo {author}
  {\bibfnamefont {C.}~\bibnamefont {Vaman}}, \bibinfo {author} {\bibfnamefont
  {K.}~\bibnamefont {Starosta}}, \ and\ \bibinfo {author} {\bibfnamefont
  {D.}~\bibnamefont {Miller}},\ }\href {\doibase DOI:
  10.1016/j.nimb.2007.04.181} {\bibfield  {journal} {\bibinfo  {journal}
  {Nuclear Instruments and Methods in Physics Research Section B}\ }\textbf
  {\bibinfo {volume} {261}},\ \bibinfo {pages} {1000 } (\bibinfo {year}
  {2007})},\ \bibinfo {note} {proceedings of the Nineteenth International
  Conference on The Application of Accelerators in Research and
  Industry}\BibitemShut {NoStop}%
\bibitem [{\citenamefont {Mueller}\ \emph {et~al.}(2001)\citenamefont
  {Mueller}, \citenamefont {Church}, \citenamefont {Glasmacher}, \citenamefont
  {Gutknecht}, \citenamefont {Hackman}, \citenamefont {Hansen}, \citenamefont
  {Hu}, \citenamefont {Miller},\ and\ \citenamefont {Quirin}}]{muelnima01}%
  \BibitemOpen
  \bibfield  {author} {\bibinfo {author} {\bibfnamefont {W.}~\bibnamefont
  {Mueller}}, \bibinfo {author} {\bibfnamefont {J.}~\bibnamefont {Church}},
  \bibinfo {author} {\bibfnamefont {T.}~\bibnamefont {Glasmacher}}, \bibinfo
  {author} {\bibfnamefont {D.}~\bibnamefont {Gutknecht}}, \bibinfo {author}
  {\bibfnamefont {G.}~\bibnamefont {Hackman}}, \bibinfo {author} {\bibfnamefont
  {P.}~\bibnamefont {Hansen}}, \bibinfo {author} {\bibfnamefont
  {Z.}~\bibnamefont {Hu}}, \bibinfo {author} {\bibfnamefont {K.}~\bibnamefont
  {Miller}}, \ and\ \bibinfo {author} {\bibfnamefont {P.}~\bibnamefont
  {Quirin}},\ }\href@noop {} {\bibfield  {journal} {\bibinfo  {journal}
  {Nuclear Instruments and Methods in Physics Research, A}\ }\textbf {\bibinfo
  {volume} {466}},\ \bibinfo {pages} {492} (\bibinfo {year}
  {2001})}\BibitemShut {NoStop}%
\bibitem [{\citenamefont {Daugas}\ \emph {et~al.}(2000)\citenamefont {Daugas},
  \citenamefont {Grzywacz}, \citenamefont {Lewitowicz}, \citenamefont
  {Achouri}, \citenamefont {Ang{\'e}lique}, \citenamefont {Baiborodin},
  \citenamefont {Bennaceur}, \citenamefont {Bentida}, \citenamefont
  {B{\'e}raud}, \citenamefont {Borcea} \emph {et~al.}}]{dauplb00}%
  \BibitemOpen
  \bibfield  {author} {\bibinfo {author} {\bibfnamefont {J.}~\bibnamefont
  {Daugas}}, \bibinfo {author} {\bibfnamefont {R.}~\bibnamefont {Grzywacz}},
  \bibinfo {author} {\bibfnamefont {M.}~\bibnamefont {Lewitowicz}}, \bibinfo
  {author} {\bibfnamefont {L.}~\bibnamefont {Achouri}}, \bibinfo {author}
  {\bibfnamefont {J.}~\bibnamefont {Ang{\'e}lique}}, \bibinfo {author}
  {\bibfnamefont {D.}~\bibnamefont {Baiborodin}}, \bibinfo {author}
  {\bibfnamefont {K.}~\bibnamefont {Bennaceur}}, \bibinfo {author}
  {\bibfnamefont {R.}~\bibnamefont {Bentida}}, \bibinfo {author} {\bibfnamefont
  {R.}~\bibnamefont {B{\'e}raud}}, \bibinfo {author} {\bibfnamefont
  {C.}~\bibnamefont {Borcea}},  \emph {et~al.},\ }\href@noop {} {\bibfield
  {journal} {\bibinfo  {journal} {Physics Letters B}\ }\textbf {\bibinfo
  {volume} {476}},\ \bibinfo {pages} {213} (\bibinfo {year}
  {2000})}\BibitemShut {NoStop}%
\bibitem [{\citenamefont {Bateman}(1910)}]{bat1910}%
  \BibitemOpen
  \bibfield  {author} {\bibinfo {author} {\bibfnamefont {H.}~\bibnamefont
  {Bateman}},\ }in\ \href@noop {} {\emph {\bibinfo {booktitle} {Proc. Cambridge
  Phil. Soc}}},\ Vol.~\bibinfo {volume} {15}\ (\bibinfo {year} {1910})\ pp.\
  \bibinfo {pages} {423--427}\BibitemShut {NoStop}%
\bibitem [{\citenamefont {Rajabali}(2009)}]{mmrthesis}%
  \BibitemOpen
  \bibfield  {author} {\bibinfo {author} {\bibfnamefont {M.~M.}\ \bibnamefont
  {Rajabali}},\ }\emph {\bibinfo {title} {Beta-decay, beta-delayed neutron
  emission and isomer studies around 78Ni}},\ \href@noop {} {Ph.D. thesis},\
  \bibinfo  {school} {University of Tennessee} (\bibinfo {year} {2009}),\
  \bibinfo {note}
  {\url{http://trace.tennessee.edu/utk_graddiss/648}}\BibitemShut {NoStop}%
\bibitem [{\citenamefont {Rajabali}(2011)}]{mmr72co}%
  \BibitemOpen
  \bibfield  {author} {\bibinfo {author} {\bibfnamefont {M.~M.}\ \bibnamefont
  {Rajabali}},\ }\href@noop {} {\enquote {\bibinfo {title} {$\beta$ decay of
  $^{72}$co and microsecond isomers in even-mass, neutron rich nickel
  isotopes},}\ }\bibinfo {howpublished} {to be published} (\bibinfo {year}
  {2011})\BibitemShut {NoStop}%
\bibitem [{\citenamefont {Audi}\ \emph {et~al.}(2003)\citenamefont {Audi},
  \citenamefont {Wapstra},\ and\ \citenamefont {Thibault}}]{audinpa03}%
  \BibitemOpen
  \bibfield  {author} {\bibinfo {author} {\bibfnamefont {G.}~\bibnamefont
  {Audi}}, \bibinfo {author} {\bibfnamefont {A.}~\bibnamefont {Wapstra}}, \
  and\ \bibinfo {author} {\bibfnamefont {C.}~\bibnamefont {Thibault}},\
  }\href@noop {} {\bibfield  {journal} {\bibinfo  {journal} {Nuclear Physics,
  Section A}\ }\textbf {\bibinfo {volume} {729}},\ \bibinfo {pages} {337}
  (\bibinfo {year} {2003})}\BibitemShut {NoStop}%
\bibitem [{\citenamefont {Rajabali}(2007)}]{mmr2007}%
  \BibitemOpen
  \bibfield  {author} {\bibinfo {author} {\bibfnamefont {M.~M.}\ \bibnamefont
  {Rajabali}},\ }in\ \href@noop {} {\emph {\bibinfo {booktitle} {Proceedings of
  the Fourth International Conference Fission and properties of neutron rich
  nuclei, Sanibel island, USA}}}\ (\bibinfo {year} {2007})\ p.\ \bibinfo
  {pages} {679}\BibitemShut {NoStop}%
\bibitem [{\citenamefont {Pawlat}\ \emph {et~al.}(1994)\citenamefont {Pawlat},
  \citenamefont {Broda}, \citenamefont {Królas}, \citenamefont {Maj},
  \citenamefont {Zieblinski}, \citenamefont {Grawe}, \citenamefont {Schubart},
  \citenamefont {Maier}, \citenamefont {Heese}, \citenamefont {Kluge},\ and\
  \citenamefont {Schramm}}]{Pawlat1994}%
  \BibitemOpen
  \bibfield  {author} {\bibinfo {author} {\bibfnamefont {T.}~\bibnamefont
  {Pawlat}}, \bibinfo {author} {\bibfnamefont {R.}~\bibnamefont {Broda}},
  \bibinfo {author} {\bibfnamefont {W.}~\bibnamefont {Królas}}, \bibinfo
  {author} {\bibfnamefont {A.}~\bibnamefont {Maj}}, \bibinfo {author}
  {\bibfnamefont {M.}~\bibnamefont {Zieblinski}}, \bibinfo {author}
  {\bibfnamefont {H.}~\bibnamefont {Grawe}}, \bibinfo {author} {\bibfnamefont
  {R.}~\bibnamefont {Schubart}}, \bibinfo {author} {\bibfnamefont {K.~H.}\
  \bibnamefont {Maier}}, \bibinfo {author} {\bibfnamefont {J.}~\bibnamefont
  {Heese}}, \bibinfo {author} {\bibfnamefont {H.}~\bibnamefont {Kluge}}, \ and\
  \bibinfo {author} {\bibfnamefont {M.}~\bibnamefont {Schramm}},\ }\href
  {\doibase DOI: 10.1016/0375-9474(94)90247-X} {\bibfield  {journal} {\bibinfo
  {journal} {Nuclear Physics A}\ }\textbf {\bibinfo {volume} {574}},\ \bibinfo
  {pages} {623 } (\bibinfo {year} {1994})}\BibitemShut {NoStop}%
\bibitem [{\citenamefont {Weissman}\ \emph {et~al.}(1999)\citenamefont
  {Weissman}, \citenamefont {Andreyev}, \citenamefont {Bruyneel}, \citenamefont
  {Franchoo}, \citenamefont {Huyse}, \citenamefont {Kruglov}, \citenamefont
  {Kudryavtsev}, \citenamefont {Mueller}, \citenamefont {Raabe}, \citenamefont
  {Reusen} \emph {et~al.}}]{weisprc99}%
  \BibitemOpen
  \bibfield  {author} {\bibinfo {author} {\bibfnamefont {L.}~\bibnamefont
  {Weissman}}, \bibinfo {author} {\bibfnamefont {A.}~\bibnamefont {Andreyev}},
  \bibinfo {author} {\bibfnamefont {B.}~\bibnamefont {Bruyneel}}, \bibinfo
  {author} {\bibfnamefont {S.}~\bibnamefont {Franchoo}}, \bibinfo {author}
  {\bibfnamefont {M.}~\bibnamefont {Huyse}}, \bibinfo {author} {\bibfnamefont
  {K.}~\bibnamefont {Kruglov}}, \bibinfo {author} {\bibfnamefont
  {Y.}~\bibnamefont {Kudryavtsev}}, \bibinfo {author} {\bibfnamefont
  {W.}~\bibnamefont {Mueller}}, \bibinfo {author} {\bibfnamefont
  {R.}~\bibnamefont {Raabe}}, \bibinfo {author} {\bibfnamefont
  {I.}~\bibnamefont {Reusen}},  \emph {et~al.},\ }\href@noop {} {\bibfield
  {journal} {\bibinfo  {journal} {Physical Review C}\ }\textbf {\bibinfo
  {volume} {59}},\ \bibinfo {pages} {2004} (\bibinfo {year}
  {1999})}\BibitemShut {NoStop}%
\bibitem [{\citenamefont {Starosta}(2009)}]{stapriv}%
  \BibitemOpen
  \bibfield  {author} {\bibinfo {author} {\bibfnamefont {K.}~\bibnamefont
  {Starosta}},\ }\href@noop {} {}\bibinfo {howpublished} {Private
  communication} (\bibinfo {year} {2009})\BibitemShut {NoStop}%
\bibitem [{\citenamefont {Hirata}(1970)}]{hirata1970}%
  \BibitemOpen
  \bibfield  {author} {\bibinfo {author} {\bibfnamefont {M.}~\bibnamefont
  {Hirata}},\ }\href@noop {} {\bibfield  {journal} {\bibinfo  {journal}
  {Physics Letters B}\ }\textbf {\bibinfo {volume} {32}},\ \bibinfo {pages}
  {656} (\bibinfo {year} {1970})}\BibitemShut {NoStop}%
\bibitem [{\citenamefont {Osnes}\ and\ \citenamefont
  {Warke}(1970)}]{osnes1970}%
  \BibitemOpen
  \bibfield  {author} {\bibinfo {author} {\bibfnamefont {E.}~\bibnamefont
  {Osnes}}\ and\ \bibinfo {author} {\bibfnamefont {C.}~\bibnamefont {Warke}},\
  }\href@noop {} {\bibfield  {journal} {\bibinfo  {journal} {Nuclear Physics,
  Section A}\ }\textbf {\bibinfo {volume} {154}},\ \bibinfo {pages} {331}
  (\bibinfo {year} {1970})}\BibitemShut {NoStop}%
\bibitem [{\citenamefont {Horoshko}\ \emph {et~al.}(1970)\citenamefont
  {Horoshko}, \citenamefont {Cline},\ and\ \citenamefont
  {Lesser}}]{horoshko1970}%
  \BibitemOpen
  \bibfield  {author} {\bibinfo {author} {\bibfnamefont {R.}~\bibnamefont
  {Horoshko}}, \bibinfo {author} {\bibfnamefont {D.}~\bibnamefont {Cline}}, \
  and\ \bibinfo {author} {\bibfnamefont {P.}~\bibnamefont {Lesser}},\
  }\href@noop {} {\bibfield  {journal} {\bibinfo  {journal} {Nuclear Physics,
  Section A}\ }\textbf {\bibinfo {volume} {149}},\ \bibinfo {pages} {562}
  (\bibinfo {year} {1970})}\BibitemShut {NoStop}%
\bibitem [{\citenamefont {Zamick}(1971)}]{zamick1971}%
  \BibitemOpen
  \bibfield  {author} {\bibinfo {author} {\bibfnamefont {L.}~\bibnamefont
  {Zamick}},\ }\href@noop {} {\bibfield  {journal} {\bibinfo  {journal}
  {Physics Letters B}\ }\textbf {\bibinfo {volume} {34}},\ \bibinfo {pages}
  {472} (\bibinfo {year} {1971})}\BibitemShut {NoStop}%
\bibitem [{\citenamefont {De~Shalit}\ and\ \citenamefont
  {Feshbach}(1974)}]{desfes74}%
  \BibitemOpen
  \bibfield  {author} {\bibinfo {author} {\bibfnamefont {A.}~\bibnamefont
  {De~Shalit}}\ and\ \bibinfo {author} {\bibfnamefont {H.}~\bibnamefont
  {Feshbach}},\ }\href@noop {} {\emph {\bibinfo {title} {{Theoretical nuclear
  physics}}}}\ (\bibinfo  {publisher} {John Wiley \& Sons},\ \bibinfo {year}
  {1974})\BibitemShut {NoStop}%
\bibitem [{\citenamefont {Franchoo}\ \emph {et~al.}(1998)\citenamefont
  {Franchoo}, \citenamefont {Huyse}, \citenamefont {Kruglov}, \citenamefont
  {Kudryavtsev}, \citenamefont {Mueller}, \citenamefont {Raabe}, \citenamefont
  {Reusen}, \citenamefont {Van~Duppen}, \citenamefont {Van~Roosbroeck},
  \citenamefont {Vermeeren}, \citenamefont {W\"ohr}, \citenamefont {Kratz},
  \citenamefont {Pfeiffer},\ and\ \citenamefont {Walters}}]{franprl98}%
  \BibitemOpen
  \bibfield  {author} {\bibinfo {author} {\bibfnamefont {S.}~\bibnamefont
  {Franchoo}}, \bibinfo {author} {\bibfnamefont {M.}~\bibnamefont {Huyse}},
  \bibinfo {author} {\bibfnamefont {K.}~\bibnamefont {Kruglov}}, \bibinfo
  {author} {\bibfnamefont {Y.}~\bibnamefont {Kudryavtsev}}, \bibinfo {author}
  {\bibfnamefont {W.~F.}\ \bibnamefont {Mueller}}, \bibinfo {author}
  {\bibfnamefont {R.}~\bibnamefont {Raabe}}, \bibinfo {author} {\bibfnamefont
  {I.}~\bibnamefont {Reusen}}, \bibinfo {author} {\bibfnamefont
  {P.}~\bibnamefont {Van~Duppen}}, \bibinfo {author} {\bibfnamefont
  {J.}~\bibnamefont {Van~Roosbroeck}}, \bibinfo {author} {\bibfnamefont
  {L.}~\bibnamefont {Vermeeren}}, \bibinfo {author} {\bibfnamefont
  {A.}~\bibnamefont {W\"ohr}}, \bibinfo {author} {\bibfnamefont {K.~L.}\
  \bibnamefont {Kratz}}, \bibinfo {author} {\bibfnamefont {B.}~\bibnamefont
  {Pfeiffer}}, \ and\ \bibinfo {author} {\bibfnamefont {W.~B.}\ \bibnamefont
  {Walters}},\ }\href {\doibase 10.1103/PhysRevLett.81.3100} {\bibfield
  {journal} {\bibinfo  {journal} {Physical Review Letters}\ }\textbf {\bibinfo
  {volume} {81}},\ \bibinfo {pages} {3100} (\bibinfo {year}
  {1998})}\BibitemShut {NoStop}%
\bibitem [{\citenamefont {Baglin}(1997)}]{Baglin1997}%
  \BibitemOpen
  \bibfield  {author} {\bibinfo {author} {\bibfnamefont {C.~M.}\ \bibnamefont
  {Baglin}},\ }\href {\doibase DOI: 10.1006/ndsh.1997.0001} {\bibfield
  {journal} {\bibinfo  {journal} {Nuclear Data Sheets}\ }\textbf {\bibinfo
  {volume} {80}},\ \bibinfo {pages} {1 } (\bibinfo {year} {1997})}\BibitemShut
  {NoStop}%
\bibitem [{\citenamefont {Machleidt}\ \emph {et~al.}(1996)\citenamefont
  {Machleidt}, \citenamefont {Sammarruca},\ and\ \citenamefont
  {Song}}]{cdbon96}%
  \BibitemOpen
  \bibfield  {author} {\bibinfo {author} {\bibfnamefont {R.}~\bibnamefont
  {Machleidt}}, \bibinfo {author} {\bibfnamefont {F.}~\bibnamefont
  {Sammarruca}}, \ and\ \bibinfo {author} {\bibfnamefont {Y.}~\bibnamefont
  {Song}},\ }\href {\doibase 10.1103/PhysRevC.53.R1483} {\bibfield  {journal}
  {\bibinfo  {journal} {Physical Review C}\ }\textbf {\bibinfo {volume} {53}},\
  \bibinfo {pages} {R1483} (\bibinfo {year} {1996})}\BibitemShut {NoStop}%
\bibitem [{\citenamefont {Machleidt}(2001)}]{cdbon01}%
  \BibitemOpen
  \bibfield  {author} {\bibinfo {author} {\bibfnamefont {R.}~\bibnamefont
  {Machleidt}},\ }\href {\doibase 10.1103/PhysRevC.63.024001} {\bibfield
  {journal} {\bibinfo  {journal} {Physical Review C}\ }\textbf {\bibinfo
  {volume} {63}},\ \bibinfo {pages} {024001} (\bibinfo {year}
  {2001})}\BibitemShut {NoStop}%
\bibitem [{\citenamefont {V.}\ and\ \citenamefont
  {Paar}(1973)}]{paar_npa_1973}%
  \BibitemOpen
  \bibfield  {author} {\bibinfo {author} {\bibnamefont {V.}}\ and\ \bibinfo
  {author} {\bibnamefont {Paar}},\ }\href {\doibase
  10.1016/0375-9474(73)90763-X} {\bibfield  {journal} {\bibinfo  {journal}
  {Nuclear Physics A}\ }\textbf {\bibinfo {volume} {211}},\ \bibinfo {pages}
  {29 } (\bibinfo {year} {1973})}\BibitemShut {NoStop}%
\bibitem [{\citenamefont {Hosmer}\ \emph {et~al.}(2010)\citenamefont {Hosmer},
  \citenamefont {Schatz}, \citenamefont {Aprahamian}, \citenamefont {Arndt},
  \citenamefont {Clement}, \citenamefont {Estrade}, \citenamefont {Farouqi},
  \citenamefont {Kratz}, \citenamefont {Liddick}, \citenamefont {Lisetskiy},
  \citenamefont {Mantica}, \citenamefont {M\"oller}, \citenamefont {Mueller},
  \citenamefont {Montes}, \citenamefont {Morton}, \citenamefont {Ouellette},
  \citenamefont {Pellegrini}, \citenamefont {Pereira}, \citenamefont
  {Pfeiffer}, \citenamefont {Reeder}, \citenamefont {Santi}, \citenamefont
  {Steiner}, \citenamefont {Stolz}, \citenamefont {Tomlin}, \citenamefont
  {Walters},\ and\ \citenamefont {W\"ohr}}]{hosprc10}%
  \BibitemOpen
  \bibfield  {author} {\bibinfo {author} {\bibfnamefont {P.}~\bibnamefont
  {Hosmer}}, \bibinfo {author} {\bibfnamefont {H.}~\bibnamefont {Schatz}},
  \bibinfo {author} {\bibfnamefont {A.}~\bibnamefont {Aprahamian}}, \bibinfo
  {author} {\bibfnamefont {O.}~\bibnamefont {Arndt}}, \bibinfo {author}
  {\bibfnamefont {R.~R.~C.}\ \bibnamefont {Clement}}, \bibinfo {author}
  {\bibfnamefont {A.}~\bibnamefont {Estrade}}, \bibinfo {author} {\bibfnamefont
  {K.}~\bibnamefont {Farouqi}}, \bibinfo {author} {\bibfnamefont {K.-L.}\
  \bibnamefont {Kratz}}, \bibinfo {author} {\bibfnamefont {S.~N.}\ \bibnamefont
  {Liddick}}, \bibinfo {author} {\bibfnamefont {A.~F.}\ \bibnamefont
  {Lisetskiy}}, \bibinfo {author} {\bibfnamefont {P.~F.}\ \bibnamefont
  {Mantica}}, \bibinfo {author} {\bibfnamefont {P.}~\bibnamefont {M\"oller}},
  \bibinfo {author} {\bibfnamefont {W.~F.}\ \bibnamefont {Mueller}}, \bibinfo
  {author} {\bibfnamefont {F.}~\bibnamefont {Montes}}, \bibinfo {author}
  {\bibfnamefont {A.~C.}\ \bibnamefont {Morton}}, \bibinfo {author}
  {\bibfnamefont {M.}~\bibnamefont {Ouellette}}, \bibinfo {author}
  {\bibfnamefont {E.}~\bibnamefont {Pellegrini}}, \bibinfo {author}
  {\bibfnamefont {J.}~\bibnamefont {Pereira}}, \bibinfo {author} {\bibfnamefont
  {B.}~\bibnamefont {Pfeiffer}}, \bibinfo {author} {\bibfnamefont
  {P.}~\bibnamefont {Reeder}}, \bibinfo {author} {\bibfnamefont
  {P.}~\bibnamefont {Santi}}, \bibinfo {author} {\bibfnamefont
  {M.}~\bibnamefont {Steiner}}, \bibinfo {author} {\bibfnamefont
  {A.}~\bibnamefont {Stolz}}, \bibinfo {author} {\bibfnamefont {B.~E.}\
  \bibnamefont {Tomlin}}, \bibinfo {author} {\bibfnamefont {W.~B.}\
  \bibnamefont {Walters}}, \ and\ \bibinfo {author} {\bibfnamefont
  {A.}~\bibnamefont {W\"ohr}},\ }\href {\doibase 10.1103/PhysRevC.82.025806}
  {\bibfield  {journal} {\bibinfo  {journal} {Phys. Rev. C}\ }\textbf {\bibinfo
  {volume} {82}},\ \bibinfo {pages} {025806} (\bibinfo {year}
  {2010})}\BibitemShut {NoStop}%
\end{thebibliography}

%

\end{document}